\documentstyle[epsf,11pt]{article}

\makeatletter

\if@twoside
\else
\fi
\ifcase \@ptsize
\or 
\or 
\fi

\renewcommand{\theequation}{\thesection.\arabic{equation}}
\@addtoreset{equation}{section}
\def\section{\@startsection {section}{1}{\z@}{-3.5ex plus -1ex minus
 -.2ex}{2.3ex plus .2ex}{\large\bf\centering}}
\def\subsection{\@startsection{subsection}{2}{\z@}{-3.25ex plus%
 -1ex minus -.2ex}{1.5ex plus .2ex}{\bf}}
\def\subsubsection{\@startsection{subsubsection}{3}{\z@}{-3.25ex plus%
 -1ex minus -.2ex}{1.5ex plus .2ex}{\sl}}
\gdef\@publabel{\hfil}
\gdef\@pubdate{\null}
\gdef\@pubnumber{\null}
\gdef\@author{\null}
\gdef\@title{\null}
\gdef\@abstract{\null}
\long\def\pubdate#1{\gdef\@pubdate{#1}}
\long\def\pubnumber#1{\gdef\@pubnumber{#1}}
\long\def\publabel#1{\gdef\@publabel{#1}}
\long\def\author#1{\gdef\@author{#1}}
\long\def\title#1{\gdef\@title{#1}}
\long\def\abstract#1{\gdef\@abstract{#1}}
\def\titlerelax{
\let\maketitle\relax
\let\settitleparameters\relax
\let\consolidatetitle\relax
\let\inittitlepage\relax
\let\finishtitlepage\relax
\let\titlepagecontents\relax
\let\multithanks\relax
\let\titlebaselines\relax
\let\@makepub\relax
\let\@maketitle\relax
\let\@makeauthor\relax
\let\@makeabstract\relax
\let\@maketitlenote\relax
\let\thanks\relax
\let\titlerelax\relax}
\def\titleclean
{\gdef\@titlenote{}
\gdef\@abstract{}
\gdef\@author{}
\gdef\@title{}
\gdef\@pubdate{}\gdef\@pubnumber{}\gdef\@publabel{}
\gdef\@dpublabel{}
}

\def\@makepub{\vbox to \z@{\hbox to \textwidth{\hfill
\@publabel \hfill
\llap{\parbox[t]{0.33\textwidth}{\raggedleft\@pubnumber}}}%
\vss}}
\def\@maketitle{\vskip 60pt \begin{center}
 {\LARGE \@title \par}
 \end{center}}

\def\@makeauthor{{%
\def\and{\smallskip {\normalsize \rm and\smallskip }}
\def\And{\medskip {\normalsize \rm and\\}\medskip}
\long\def\address##1{{\def\and{\\and\\}\medskip
                                {\small \it \\##1\\}
}}
{\centering
 \vskip 3em
 \large \lineskip .75em
 \@author}
 \par}}
\def\@makedate{\vskip 1.5em
 {\raggedright \small \noindent\@pubdate \par}}
\def\@makeabstract{\vskip 1.5em
{\small
\begin{center}
{\bf ABSTRACT\vspace{-.5em}\vspace{0pt}}
\end{center}
\quotation \@abstract \endquotation}}
\def\maketitle{\titlepage
\let\footnotesize\small \setcounter{page}{0}
\@makepub
\vfil
\@maketitle
\@makeauthor
\vfil
\@makeabstract
\@thanks
\vfil
\@makedate
\if@restonecol\twocolumn \else \eject \fi
\titlerelax \titleclean
\setcounter{footnote}{0}
}

\newcommand{\ncm}{\newcommand}
\ncm{\ol}{\overline}
\ncm{\ul}{\underline}
\ncm{\ve}{\varepsilon}
\ncm{\gf}{\bar{g}}
\ncm{\hf}{\bar{h}}
\ncm{\Wf}{\ol{W}}
\ncm{\ga}{g}
\ncm{\si}{\sigma}
\ncm{\h}{h^{\vee}}
\ncm{\kh}{k+\h}
\ncm{\alphav}{\alpha^{\vee}}
\ncm{\lambdab}{\bar{\lambda}}
\ncm{\rhob}{\bar{\rho}}
\ncm{\mub}{\bar{\mu}}

\ncm{\xmax}{\ol{x}}
\ncm{\xmin}{\ul{x}}
\ncm{\ymax}{\ol{y}}
\ncm{\ymin}{\ul{y}}
\ncm{\wmax}{\ol{w}}
\ncm{\wmin}{\ul{w}}
\ncm{\zmax}{\ol{z}}
\ncm{\zmin}{\ul{z}}

\ncm{\ha}{h}
\ncm{\na}{n}
\ncm{\Da}{\Delta}
\ncm{\Qa}{Q}
\ncm{\Qf}{\ol{Q}}
\ncm{\Wa}{W}

\ncm{\ch}{\mbox{ch}\,}
\ncm{\W}{{\cal W}}

\ncm{\la}{\langle}
\ncm{\ra}{\rangle}

\newtheorem{conjecture}{Conjecture}

\ncm{\be}{\begin{equation}
\addtolength{\abovedisplayskip}{\extraspaces}
\addtolength{\belowdisplayskip}{\extraspaces}
\addtolength{\abovedisplayshortskip}{\extraspace}
\addtolength{\belowdisplayshortskip}{\extraspace}}
\ncm{\ee}{\end{equation}}
\ncm{\bea}{\begin{eqnarray}
\addtolength{\abovedisplayskip}{\extraspaces}
\addtolength{\belowdisplayskip}{\extraspaces}
\addtolength{\abovedisplayshortskip}{\extraspace}
\addtolength{\belowdisplayshortskip}{\extraspace}}
\ncm{\eea}{\end{eqnarray}}
\ncm{\beas}{\begin{eqnarray*}
\addtolength{\abovedisplayskip}{\extraspaces}
\addtolength{\belowdisplayskip}{\extraspaces}
\addtolength{\abovedisplayshortskip}{\extraspace}
\addtolength{\belowdisplayshortskip}{\extraspace}}
\ncm{\eeas}{\end{eqnarray*}}
\newcommand{\ie}{{\it i.e}.\ }
\newcommand{\eg}{{\it e.g}.\ }
\newlength{\extraspace}
\newlength{\extraspaces}
\begin{document}
\pubnumber{BONN-TH-95-14 \\ULB-TH-95/09}
\pubdate{}
\title{The Kazhdan-Lusztig conjecture for $\W$-algebras}

\author{KOOS DE VOS
\address{
Physikalisches Institut der Universit\"at Bonn\\ Nussallee 12, 53115 Bonn, Germany}
\And
PETER VAN DRIEL
\address{
Service de Physique Th\'eorique\\
Universit\'e Libre de Bruxelles, Campus Plaine, C.\ P.\ 225\\ 
Boulevard du Triomphe, B-1050 Bruxelles, Belgium}
}

\abstract{
The main result in this paper is the character formula for arbitrary
irreducible highest weight modules of $\W$ algebras. The key
ingredient is the functor provided by quantum Hamiltonian reduction,
that constructs the $\W$ algebras from affine Kac-Moody algebras and
in a similar fashion $\W$
modules from KM modules. Assuming certain properties of this functor,
the $\W$ characters are subsequently derived from the Kazhdan-Lusztig
conjecture for KM algebras. The result can be formulated in terms of a
double coset of the Weyl group of the KM algebra: the Hasse diagrams
give the embedding diagrams of the Verma modules and the
Kazhdan-Lusztig polynomials give the multiplicities in the characters.  
}
\maketitle
%
%

\section{Introduction}
$\W$ algebras were introduced more than a decade ago as (higher spin)
extensions of the Virasoro algebra in the context of two dimensional
conformal field theory~\cite{Zam}. Analogous to the Virasoro algebra, one
expects that the representation theory of $\W$ algebras plays a
crucial role in applications such as in conformal field theories with
$\W$ symmetry, and in theories where the $\W$ symmetry is gauged ($\W$
strings and $\W$ gravity), see~\cite{BS,PH} for reviews. For these applications,
the relevant representations are highest weight modules. A basic goal
is therefore to describe the irreducible modules, and more
specifically to compute their characters.  \\
There exists a general approach to find the irreducible  characters from
the characters of Verma modules. Any Verma module $M(x)$ can be
decomposed into irreducible highest weight modules $L(y)$ (local
composition series). This gives rise to character formulas of the form
\be\label{eq:ML}
\ch M(x)  = \sum_y m_{xy} \, \ch L(y),
\ee
where $m$ is a matrix whose entries $m_{xy}$ count the number of times
that $L(y)$ appears in the decomposition of $M(x)$. Doing this for all
$M(x)$ such that $m$ can be inverted gives
\be\label{eq:LM}
\ch L(x)  = \sum_y m^{-1}_{xy} \, \ch M(y).
\ee
The characters of Verma modules are in general easy to compute, hence the
computation of the characters of the irreducible modules boils down to
determining the multiplicities $m_{xy}$.  \\
This general programme has been applied successfully to
the Virasoro algebra~\cite{FF}. The key ingredient there is that
every submodule of a Verma module is a sum of Verma modules.
Since there is at most one embedding between Verma modules this implies
that the multiplicities $m_{xy}$ are 0 or 1, and the irreducible
characters follow directly from the embedding pattern of the Verma
modules. These embedding patterns are completely classified, and
consequently for the Virasoro algebra, the characters of {\em all}
irreducible highest weight modules are known.   \\ 
For $\W$ algebras the submodule structure of Verma modules is much
more complicated: in general submodules are not sums of Verma modules.
Therefore the embedding patterns of the Verma modules do not determine
the irreducible characters. This is directly related to the occurence
of multiplicities $m_{xy}\!>\!1$.\\  
There are of course also other approaches. For instance, for the
$\W_N$ minimal models, the irreducible character $\ch L(x)$ (for $x$
inside the Kac-table) has been determined directly, using free field
methods~\cite{FL}. In terms of the multiplicities this amounts to
having computed a single row of $m^{-1}$. It does not appear to be
possible to apply these methods to compute the other rows, which is
necessary to determine the characters of all irreducible highest
weight modules (\ie also for $x$ outside or on the boundary of the
Kac-table). In a way, the results of~\cite{FL} for $\W_N$ algebras
amount to having the $\W$ analogue of the Weyl-Kac character formula
for affine Kac-Moody algebras.\\ 
For affine KM algebras the characters are known beyond the Weyl-Kac
character formula. For $\kh\neq 0$ the programme described above
has been fully completed. The result can be summarized as follows:  \\ 
(1) The weights $y$ appearing in the decomposition~(\ref{eq:ML}) are
determined by a subgroup of the affine Weyl group, and the
associated Bruhat ordering (the Kac-Kazhdan condition~\cite{KK})\\
(2) The multiplicities $m_{xy}$ are given in terms of the
Kazhdan-Lusztig polynomials associated to the affine Weyl group (the
Kazhdan-Lusztig conjecture~\cite{KL1,DGK})\\
The main ingredient in the proof of (1) is the Jantzen filtration,
whereas (2) has been proven using the intersection cohomology of Schubert
varieties (only for integral weights, for other weights it is still a
conjecture). Neither of these concepts seems to have been worked out
for $\W$ algebras. \\ 
It is now interesting to note that $\W$ algebras and KM algebras are
intimately related. In particular, a large class of $\W$ algebras can
be obtained from affine KM algebras by (quantum) Hamiltonian
reduction, where one imposes certain constraints on the KM
generators~(see \cite{Fea} for a review).  In this way a $\W$ algebra
can be constructed for every embedding of $sl_2$ into the simple Lie
algebra underlying the affine KM algebra~\cite{BTV}.  The quantum
construction naturally allows for a BRST formulation, in which the
$\W$ algebra arises as the BRST cohomology of a complex involving the
KM algebra~\cite{FeFr,FKW,BT}. Of course, given an $sl_2$ embedding, one can also compute the
cohomology of a KM module. By construction, the result will be a
module of the corresponding $\W$ algebra. Thus, one obtains in a
natural way a functor from KM modules to $\W$ modules. The
action of this `reduction functor' is in general hard to compute.
In~\cite{FKW}, the action on (resolutions of) admissible KM modules
was computed for principal $sl_2$ embeddings, assuming certain
properties of the reduction functor. This way the characters of the
$\W_N$ minimal models are recovered.  \\
The main new idea in this paper is to apply the reduction functor to
`arbitrary' KM modules, to find the analogues of the general results (1)
and (2) for $\W$ algebras. The result is a natural generalization of
the KL conjecture to $\W$ algebras associated to arbitrary $sl_2$
embeddings. We show how this `KL conjecture for $\W$
algebras' can be derived from the KL conjecture for KM algebras,
assuming similar properties as in~\cite{FKW} of the reduction functor.
These assumptions are motivated by the results~\cite{DV}
for finite $\W$ algebras. The upshot is that {\em all} irreducible
characters for such $\W$ algebras are thereby determined. We verified
the conjecture for a nontrivial set of $\W_3$ modules. \\ 
The setup of this paper is as follows. In section 2, we review the
representation theory and KL conjectures of affine KM algebras,
including a discussion of the translation functor that serves as a
helpful analogy with the reduction functor. Then in section 3, after
some remarks on the representation theory of general $\W$ algebras, we
present the main result of this paper in section 3.2, the KL
conjecture for $\W$ algebras. We also give an idea of how it can be
derived using the reduction functor. Several applications of the conjecture
are discussed in section 4. Properties of Coxeter groups and their KL
polynomials are given in the appendix.  
\section{The Kazhdan-Lusztig conjectures for affine Kac-Moody algebras}
We first present a collection of results concerning affine KM algebras
and their highest weight modules, leading to the KL
conjectures. Virtually everything stated here can be found somewhere
in the mathematical literature on the subject, or can be concluded
directly from it. We have avoided a rigorous presentation, but instead
focussed on the line of thought, and made clear what is
well-established and what is conjectured. 
For background and explanations on KM algebras and the structure of
the highest weight modules we refer to~\cite{Jant,Kac}, and
for Weyl groups and KL polynomials to~\cite{Hu}. 
\subsection{Composition series and character formulae}
Let $\ga$ be an affine KM algebra, and fix a triangular
decomposition $\ga=\na_+\oplus \ha\oplus \na_-$ in positive root
generators, Cartan subalgebra and negative root generators. A singular
vector $v_{\lambda}$ is an eigenvector of the generators of the Cartan
subalgebra $\ha$ with weight $\lambda\in\ha^*$, and is annihilated by
the positive root generators. A highest weight module is a module that is generated from
a singular vector, the highest weight vector, by the action of the negative root generators.
There are two important examples of highest weight modules. The first
is the Verma module $M(\lambda)$, which is uniquely defined by the
property that it is generated freely from $v_{\lambda}$. The second is the quotient of
$M(\lambda)$ by its maximal proper submodule, which gives the unique
irreducible highest weight module $L(\lambda)$. \\  
Highest weight modules themselves are special examples of modules in the
so-called category ${\cal O}$~\cite{BGG,DGK}. In general this category consists
of modules $V$ which have a weight space decomposition
\be
\label{eq:weightdeco}
V = \oplus_{\mu \leq \lambda} V_{\mu} \;,
\ee
where the $\mu$'s satisfy $\mu \leq \lambda$ for $\lambda$ in some
finite subset of $h^*$ (recall that $\mu\leq\lambda$ iff
$\lambda-\mu$ is on the positive root lattice $Q_+$ of $\ga$) and $\dim
V_{\mu}<\infty$. The category ${\cal O}$ contains highest weight
modules, tensor products, submodules, quotients, etc.\\
For every module $V$ in ${\cal O}$, one can define a (formal) character 
$\ch V$,  
\be\label{eq:charv}
\ch V = \sum_{\mu} \dim V_{\mu} \: e^{\mu} \;,
\ee
where the formal exponentials satisfy $e^{\lambda} e^{\mu} =
e^{\lambda + \mu}$ and $e^0=1$. The character of the Verma module
$M(\lambda)$ is given by
\be
\label{eq:vchar}
\ch M(\lambda) = e^{\lambda} \sum_{\gamma \in \Qa_+}  P(\gamma)
e^{-\gamma} = e^{\lambda} \prod_{\alpha \in \Da_+}
(1-e^{-\alpha})^{-\dim \ga_{\alpha}} \;,
\ee
where $\ga_{\alpha}$ is the root space of root $\alpha$, $\Delta_+$ is
the set of positive roots and $P(\gamma)$ is the (generalized) Kostant
partition function.  
One of the central problems of representation theory is to find the characters of the
irreducible highest weight modules $L(\lambda)$. The strategy is to relate
these to the explicit characters of Verma modules~(\ref{eq:vchar}). This
is possible due to the following general structure theorem, which
also illustrates that ${\cal O}$ is natural in the context of highest
weight modules (in particular, the $L(\lambda)$'s are the only
irreducibles in ${\cal O}$). Every module $V$ in the category
$\cal{O}$ has a local composition series at any weight $\lambda$ of
$V$. A local composition series for $V$ at $\lambda$ is a sequence of submodules of
$V$, $V=V_0 \supset V_1 \supset \ldots \supset V_{n-1} \supset V_n =
0$, such that either $V_i/V_{i+1} \cong L(\mu)$ for some $\mu
\geq \lambda$, or $(V_i/V_{i+1})_{\mu} = 0$ for all $\mu \geq
\lambda$. 
One denotes by $[V\!:\!L(\mu)]$ the number of
times that $L(\mu)$ appears in the local composition series of $V$ at
$\lambda$, it is called the multiplicity of $L(\mu)$ in $V$. It is
independent of the particular sequence of submodules one chooses. We stress
that $[V\!:\!L(\mu)]$ does not count the number of singular vectors at
weight $\mu$ in $V$: the statement that $V_i/V_{i+1}
\cong L(\mu)$ only requires that there is a vector $v_{\mu}$ which is
singular in the quotient $V_i/V_{i+1}$  but not necessarily singular
in $V_i$, let alone $V$. A vector that is singular in a quotient of
submodules is called primitive, and the corresponding weight is called
a primitive weight. Obviously, a singular vector is also primitive,
but it is important to realize that there are also other types of
primitive vectors. We also stress that the multiplicities $[M:L]$ can
be larger than $1$, contrary to what was initially thought based on
the known trivial multiplicities of the simple Lie algebras
$\gf=\ol{sl}_2,\ol{sl}_3$ and the affine KM algebra $g=sl_2$. 
\\  
At the level of characters, the local composition series implies
that~(\ref{eq:charv}) is given by a sum over the irreducible
characters: $\ch V = \sum_{\mu} [V\!:\!L(\mu)] \: \ch L(\mu)$ where
the sum runs over the weights of $V$ (of course, only the primitive
weights give a non-vanishing contribution). This applies in particular
to Verma modules, leading to 
\be 
\label{eq:verdeco}
\ch M(\lambda) = 
\sum_{\mu\leq\lambda} [M(\lambda)\!:\!L(\mu)] \: \ch L(\mu)\;.
\ee
Note that the composition series starts with
$[M(\lambda)\!:\!L(\lambda)]=1$, since $\dim M(\lambda)_{\lambda}=1$.
Ordering the set of weights $\mu \leq \lambda$ as
$\lambda=\mu_0$, $\mu_1$, $\mu_2$, \ldots such that $j \geq i$
whenever $\mu_j \leq \mu_i$, one has the
following set of equations,
$$
\ch M(\mu_i) = 
\sum_{\mu_j \leq \mu_i} [M(\mu_i)\!:\!L(\mu_j)] \: \ch L(\mu_j)\;.
$$
The matrix $[M(\mu_i)\!:\!L(\mu_j)]$, called the Jantzen matrix (for
$\lambda$), is upper triangular with ones on the main diagonal.
Therefore, it can be inverted. Denoting the inverse matrix
elements by $(L(\mu_i):M(\mu_j))$ (which are possibly negative
integers), one finds 
$$
\ch L(\mu_i) = \sum_{\mu_j \leq \mu_i} (L(\mu_i):M(\mu_j)) \: \ch M(\mu_j).
$$
In conclusion, from~(\ref{eq:verdeco}) one finds the character formula 
\be
\label{eq:irchar}
\ch L(\lambda) = 
\sum_{\mu \leq \lambda} (L(\lambda)\!:\!M(\mu)) \: \ch M(\mu) \;.
\ee
Here $\ch M(\mu)$ is given through~(\ref{eq:vchar}). Computing $\ch
L(\lambda)$ boils down to computing the numbers
$(L(\lambda)\!:\!M(\mu))$ for all $\mu \leq \lambda$, or equivalently,
the Jantzen matrix $[M(\mu_i)\!:\!L(\mu_j)]$ for $\lambda$. 

\subsection{The Kac-Kazhdan conditions}
The first step in determining the multiplicities $[M(\lambda)\!:\!L(\mu)]$
is to find all pairs $\lambda,\mu$ such that $[M(\lambda)\!:\!L(\mu)] \neq
0$. The general solution to this problem has been given by Kac and
Kazhdan~\cite{KK}, using the generalized Casimir of $\ga$ and the
Jantzen-filtration of Verma modules~\cite{Jant}. For the
purposes of this paper it is sufficient to consider only weights
$\lambda$ with $\la\lambda+\rho,\delta\ra=\kh\neq 0$. In that
case the result of~\cite{KK} can be rephrased in terms of properties of the
affine Weyl group~\cite{Kum}.\\
The affine Weyl group $W$ is a Coxeter group, generated by the simple
reflections $s_i$ where $s_i(\lambda)=\lambda-\la\lambda,\alphav_i \ra
\alpha_i$ are the reflections in the simple roots $\alpha_i$ of $\ga$.
Arbitrary elements $w\in W$ correspond to expressions
$w=s_{i_1}s_{i_2}\ldots s_{i_k}$. The minimal number of simple
reflections needed to generate $w$ is called the length $\ell(w)$ of
$w$. An expression of minimal length is called reduced. If $w,w' \in
W$ are two reduced expressions, then we denote $w<w'$ if the reduced
expression for $w$ can be obtained by dropping simple reflections from
a reduced expression for $w'$. The resulting relation $w\leq w'$ is a
partial ordering of $W$, called the Bruhat ordering.  

An important ingredient in what follows is the subgroup
$\Wa_{\lambda}\subset\Wa$: it is the  group generated by reflections
$r_{\hat{\alpha}}$ with $\hat{\alpha} \in \Delta_{\lambda,+}^{re} = \{
\alpha \in \Delta_+^{re} | \la\lambda + \rho, \alphav \ra \in {\bf Z}
\}$. Clearly, only if $\lambda$ is integral, $W_{\lambda}=W$ (recall
that $\rho\in h^*$ satisfies $\la\rho,\alphav_i\ra=1$), otherwise
$W_{\lambda}$ will be a proper subgroup of $W$ (which for affine $W$
may be isomorphic to $W$). It can be shown that $\Wa_{\lambda}$ is
again a Weyl group, it is generated by simple reflections
$\hat{s}_i=r_{\hat{\alpha}_i}$ (simple in $W_{\lambda}$) where
$\hat{\alpha}_i$ are the simple roots of the rootsystem
$\Delta_{\lambda,+}^{re}$. The length function on $W_{\lambda}$ is
denoted $\ell_{\lambda}(w)$.  Obviously, the relation between
$\lambda$ and $\Wa_{\lambda}$ is many-to-one, for instance
$\Wa_{\lambda}=\Wa_{\lambda+\mu}$ for arbitrary integral weight $\mu$.
In fact, up to isomorphisms, there is only a finite number of
$\Wa_{\lambda}$~\cite{Dy}. \\ 
The groups $\Wa_{\lambda}$ organize the non-vanishing multiplicities
in the following way: the primitive weights of $M(\lambda)$ are
on the shifted Weyl orbit $\Wa_{\lambda}.\lambda$, where 
\be\label{eq:shift}
w.\lambda\equiv w(\lambda+\rho)-\rho,
\ee
and vice versa: only the lower weights (with respect to Bruhat
ordering) on the orbit are primitive weights of $M(\lambda)$. In the
remainder of this section we describe this in more detail.

First consider $\kh>0$. Then every orbit $\Wa_{\mu}.\mu$ has
precisely one maximal element $\lambda$, the dominant weight, such that
$w.\lambda \leq \lambda$ for all $w \in \Wa_{\mu}$. Using~(\ref{eq:shift})
it is easy to see that such a dominant weight $\lambda$ is characterised by 
\be
\la \lambda + \rho, \hat{\alpha}^{\vee}_i \ra \geq 0.
\ee 
Clearly, there is a one-to-one correspondence between dominant weights
$\lambda$ and orbits $W_{\lambda}.\lambda$.  
There may not be a one-to-one correspondence between
elements of $\Wa_{\lambda}$ and the weights on the orbit
$\Wa_{\lambda}.\lambda$. This happens precisely if there is a subgroup
$W_{\lambda}^0$ of $\Wa_{\lambda}$ which leaves $\lambda$ invariant.
$W_{\lambda}^0$ is a finite parabolic subgroup of $W_{\lambda}$, generated by the simple reflections
$r_{\hat{\alpha}_i}$ with $\hat{\alpha}_i$ satisfying  $\la\lambda
  + \rho, \hat{\alpha}^{\vee}_i \ra = 0$. 
A dominant weight is called regular if $W_{\lambda}^0$ is trivial, and
it is called singular otherwise. Thus, weights on the orbit $\Wa_{\lambda}.\lambda$ of a dominant
weight are in one-to-one correpondence with elements of the coset
\be\label{eq:scos}
\Wa_{\lambda}/W^0_{\lambda},
\ee
\ie any weight $\mu$ can be written uniquely as $\mu=w.\lambda$
with $\lambda$ dominant and $w\in\Wa_{\lambda}/W^0_{\lambda}$. This
coset will be crucial in what follows: in particular {\it the multiplicities depend on
$\lambda$ only through the coset $\Wa_{\lambda}/W^0_{\lambda}$!}\\ 
Denote $M_w=M(w.\lambda)$ and $L_w=L(w.\lambda)$, then the Kac-Kazhdan
condition for $\kh>0$ can be described as follows 
\be\label{eq:kkbruhp}
[M_w\!:\!L_{w'}] \neq 0 \quad \mbox{iff} \quad w \leq w'
\quad \mbox{with $w,w' \in \Wa_{\lambda}/W^0_{\lambda}$}. 
\ee
Here, the ordering on the coset $\Wa_{\lambda}/W^0_{\lambda}$ is induced
from the Bruhat ordering on $\Wa_{\lambda}$: 
\be
w\leq w'\quad \mbox{with}\, w,w'\in \Wa_{\lambda}/W^0_{\lambda}\qquad \mbox{iff}\qquad
\ul{w}\leq\ul{w}'\quad \mbox{with} \,\ul{w},\ul{w}'\in \Wa_{\lambda}.
\ee
(here $\ul{w}$ is the minimal coset representative of $w$ in the
coset, defined through $\ell(\ul{w}s)>\ell(\ul{w})$ for all $s\in W^0_{\lambda}$.
Of course we could also have chosen the maximal representatives $\ol{w}$ which have
$\ell(\ol{w}s)<\ell(\ol{w})$ for all $s\in W^0_{\lambda}$).   

For the character formulas~(\ref{eq:verdeco}) and~(\ref{eq:irchar})
the Kac-Kazhdan result implies the following. First of all, the sum
over the weight space in~(\ref{eq:verdeco}) reduces to a sum over
$w'\in\Wa_{\lambda}/\Wa_{\lambda}^0$  
\be \label{eq:verdeco1}
\ch M_w = 
\sum_{w'\geq w} [M_w\!:\!L_{w'}] \: \ch L_{w'}.
\ee
Secondly, using transitivity of the Bruhat order this can be inverted 
\be \label{eq:irchar1}
\ch L_w = 
\sum_{w'\geq w} (L_w\!:\!M_{w'}) \: \ch M_{w'}.
\ee
Unlike the sum in~(\ref{eq:verdeco1}) not all terms in this sum have to be
nonvanishing.

For weights with $\kh<0$ the result can be rephrased analogously. We
note that weights with $\kh<0$ are the image of weights with $\kh>0$
under the shifted inversion 
\be
\si.\lambda = -\lambda - 2\rho.
\ee
Clearly, $\Wa_{\si.\lambda} = \Wa_{\lambda}$, so $\si$ is also a one-to-one map
between the orbits on either side (orbits always belong to one side only
as $\Wa$ leaves $\kh$ invariant).
Since $\si$ reverses the order of weights, every orbit now will have a
minimal weight, called anti-dominant, which is of the form $\si.\lambda$
with $\lambda$ dominant. In terms of these anti-dominant weights
one has the analogue of~(\ref{eq:kkbruhp}) describing the
full KK condition for $\kh<0$ 
\be\label{eq:kkbruhn}
[M_w\!:\!L_{w'}] \neq 0 \quad \mbox{iff} \quad w \geq
w' \quad \mbox{with $w,w' \in \Wa_{\lambda}/W^0_{\lambda}$}. 
\ee
Thus one finds the same character formulas~(\ref{eq:verdeco1})
and~(\ref{eq:irchar1}) but with the sum over $w'\leq w$.

\subsection{Embeddings of Verma modules}
In the previous section we have discussed the role of the cosets
$\Wa_{\lambda}/W^0_{\lambda}$ in finding the primitive weights of a
Verma module $M(\lambda)$. In this section we discuss how the same cosets
also describe the embeddings between Verma modules. \\
This is based on the property of KM Verma modules that at every primitive
weight there is at least one singular vector~\cite{KK}. Since a singular
vector $v_{\mu}$ in a Verma module $M(\lambda)$ give rise to a
homomorphism $M(\mu) \hookrightarrow M(\lambda)$ (embedding)
between Verma modules, this statement implies that there is a
homomorphism iff the multiplicity $[M(\lambda)\!:\!L(\mu)]$ is
nonvanishing. Hence
\be\label{eq:emb}
M_{w'} \hookrightarrow  M_w \quad \mbox{iff} \quad w \leq w' \quad \mbox{with $w,w' \in \Wa_{\lambda}/W^0_{\lambda}$}. 
\ee 
In other words: the diagram representing the embeddings of the Verma
modules is given by the Hasse diagram of the coset
$\Wa_{\lambda}/W_{\lambda}^0$: the vertices of this diagram are the
elements of the coset and the links between the vertices connect the
adjacent elements (two coset elements $x,y$ are called adjacent if
there is no third coset element $z$ such that $x<z<y$). Since one can
classify the Hasse diagrams, this gives a classification of embedding
diagrams. 

In fact, if $\kh\neq 0$ the relation between embeddings and the Hasse
diagram is even stronger, because in that case there is at most 1
singular vector at every primitive weight. This implies that the
homomorphism $M(\mu) \hookrightarrow M(\lambda)$ is unique, or
\be\label{eq:dimhom}
\dim\mbox{Hom}(M(\mu),M(\lambda))\leq 1.
\ee
This can be argued as follows. If there is a sequence
$M(\mu_1)\hookrightarrow M(\mu_2)\hookrightarrow M(\mu_3)$ of
homomorphisms, the embedding property implies that 
\be\label{eq:mulincr}
\dim\mbox{Hom}(M(\mu_1),M(\mu_3))\geq\dim\mbox{Hom}(M(\mu_2),M(\mu_3)).
\ee
For $\kh<0$ any Verma module contains always a lowest primitive
weight (the anti-dominant weight). At this weight, there is precisely one
singular vector (because any two embedded Verma modules 
necessarily overlap). This immediately implies~(\ref{eq:dimhom}). \\
For $\kh>0$ there is no lowest primitive weight. In that
case~(\ref{eq:dimhom}) follows from the result for $\kh<0$ through the
`reflection principle' of semi-infinite homology~\cite{FLZ}, 
\be
\mbox{Hom}(M(\mu),M(\lambda)) \simeq \mbox{Hom}(M(\sigma.\lambda),M(\sigma.\mu)).
\ee
\subsection{Jantzens translation functor}
In this section we discuss how the multiplicities
for {\sl arbitrary} dominant weights follow from
the multiplicities for {\sl regular} dominant weights. The idea is to use Jantzens translation
functor~\cite{Jant,DGK} to map modules with regular weights (trivial
$\Wa^0_{\lambda}$) to modules with singular weights (nontrivial
$\Wa^0_{\lambda}$). The reason for highlighting this ingredient here
is the striking similarity between this derivation and the derivation of $\W$
multiplicities from KM multiplicities using the reduction functor in
section 3.2. \\ 
Let $\lambda'$ be a singular dominant weight, and let $\lambda$ be a
regular dominant weight such that $\lambda-\lambda'$ is an integral
weight. Clearly, $W_{\lambda}=W_{\lambda'}$, but $W_{\lambda}^0$ is
trivial whereas $W_{\lambda'}^0$ is not. The tensorproduct with the
irreducible module associated with $\lambda'-\lambda$ gives rise to an
exact functor~\cite{Jant,DGK,KW1} (the translation functor) that maps
\be\label{eq:map1}
M(w.\lambda)\stackrel{t}{\mapsto} M(w.\lambda').
\ee
To obtain the action of the translation functor on irreducible
modules, observe that for Verma modules $M(w'.\lambda)
\hookrightarrow M(w.\lambda)$ with $w,w'$ in the same coset, the
functor maps the quotient $M(w.\lambda)\backslash M(w'.\lambda)$
(which contains $L(w.\lambda')$) to zero, so it immediately follows that 
\be\label{eq:map2}
L(w.\lambda)\stackrel{t}{\mapsto} L(w.\lambda')\delta_{\wmax,w}.
\ee
(with $\wmax$ is the maximal representative of $w$ in the coset
$W_{\lambda}/W_{\lambda'}^0$). The maps~(\ref{eq:map1})
and~(\ref{eq:map2}) determine the multiplicities for singular weights
from the multiplicities of the regular weights:
\be
[M(w.\lambda'):L(w'.\lambda')]=[M(\wmax.\lambda):L(\wmax'.\lambda)].
\ee
Another useful application of the translation functor is the
computation of the character of the irreducible module $L_e$ for
regular dominant weights $\lambda$ (without having to determine the
full Jantzen matrix). For such weights namely, the sum
in~(\ref{eq:irchar1}) runs over all the elements of
$W_{\lambda}$. Applying~(\ref{eq:map1}) and~(\ref{eq:map2}) to it
for a translation chosen such that $W_{\lambda'}^0$ contains just one
reflection, gives that the coefficients are given by
$\ve_w=(-1)^{\ell_{\lambda}(w)}$~\cite{KW1}, hence   
\be\label{eq:kwchar}
\ch L_e = \sum_{w \in W_{\lambda}} \ve_w \ch M_w.
\ee
This is the generalisation of the Weyl-Kac formula~\cite{Kac} to
arbitrary regular dominant weights. The same trick cannot be applied to
obtain arbitrary characters (\ie $\ch L_w$ or for $\lambda$ singular).
It is this particular character formula (for admissible $\lambda$)
that forms the starting point of~\cite{FKW} for generalization to $\W$ algebras.
\subsection{The KL conjectures}
\label{kl}
Now we are ready to describe the final step, \ie to give the
Kahdan-Lusztig formula for the multiplicities. In~\cite{KL1}, Kazhdan
and Lusztig defined a set of polynomials $P_{x,y}(q)$, labelled by
pairs of elements $x,y$ for an arbitrary Coxeter group $W$, and
depending on a single variable $q$. For details and properties about
the definition of these polynomials see the appendix, important for us
is that they can be computed explicitly from a recursion relation
(see~(\ref{eq:recrel}))  
\be\label{eq:recrel2}
P_{x,ys}=q^{1-c}P_{xs,y}+q^cP_{x,y}-q\sum_{\stackrel{x\leq
    z<y}{zs<z}} P_{x,z}\slash{\!\!\!\!P}_{z,y}.
\ee
The simple reflection $s$ is chosen such that $y<ys$, such that
the polynomials $P_{x,y}$ are expressed in terms of polynomials
$P_{x',y'}$ with $\ell(y')<\ell(y)$. \\  
Similarly, one defines a set of inverse polynomials $Q_{x,y}(q)$
through
\be
\sum_{x \leq z \leq w} 
P_{x,z}(q) Q_{z,y}(q)\ve_z\ve_y = \delta_{x,y}
\ee
which can also be computed directly from a recursion relation (see~(\ref{eq:qrecrel}))
\be\label{eq:qrecrel2}
Q_{x,ys} = c Q_{xs,y} + (-q)^c Q_{x,y} + c q \sum_{\stackrel{x <
    z \leq y}{zs>z}} \slash{\!\!\!\!Q}_{x,z} Q_{z,y}
\ee
Analogously, one may also associate KL polynomials $P^I,Q^I$ to a
coset $W/W_I$ for $W_I$ a parabolic subgroup of $W$. If $W_I$ is
finite these are related to the KL polynomials on $W$ as follows
\be
\label{eq:oneside}
P^{I}_{x,y} = P_{\xmax,\ymax}, \qquad Q^{I}_{x,y} = Q_{\xmin,\ymin} 
\ee
Here $\zmin$ and $\zmax$ are the minimal and maximal representatives
of $z$ in the coset $[z]$. In general, the polynomials $P^I$ and $Q^I$
are not each others inverse. The inverse polynomials of $P^I,Q^I$ 
are denoted $\tilde{Q}^I,\tilde{P}^I$, they are defined through
\be
\sum_{x \leq z \leq y} \tilde{Q}^{I}_{x,z} P^{I}_{z,y}  =  \sum_{x \leq
  z \leq y} Q^{I}_{x,z} \tilde{P}^{I}_{z,y} = \delta_{x,y}.
\ee
They can also be expressed in terms of the polynomials on $W$
\be
\tilde{P}^{I}_{x,y} = \sum_{z \in [x]} P_{z,\ymin} \ve_z
\ve_{\ymin},\qquad  \tilde{Q}^{I}_{x,y} = \sum_{z \in [y]} Q_{\xmax,z}
\ve_{\xmax} \ve_z 
\ee
The KL conjectures relate the multiplicities in the character formulas to
the value of these polynomials at $q=1$. Let $\lambda$ be a dominant
weight with coset $\Wa_{\lambda}/\Wa_{\lambda}^0$, $P_{w,w'}$ the KL
polynomials for $W_{\lambda}$ and $Q_{w,w'}$ the associated inverse KL
polynomials. Then the multiplicities are given by~\cite{KL1,DGK,Jant,Lus1}   
\be
\label{eq:klkmd}
\begin{array}{lll}
\kh>0 : \qquad&
[M_w\!:\!L_{w'}] = P^I_{w,w'}(1), \quad&
(L_w\!:\!M_{w'}) = \tilde{Q}^I_{w,w'}(1) \\ {}
\kh<0 :\qquad &
[M_w\!:\!L_{w'}] = Q^I_{w',w}(1), \quad&
(L_w\!:\!M_{w'}) = \tilde{P}^I_{w',w}(1) 
\end{array}
\ee
(the superscript $I$ refers to the subgroup $W^0_{\lambda}$)
These conjectures have been proven for integral weights,
in~\cite{iwp} for $\kh>0$, and~\cite{iwn} for $\kh<0$. It is not
inconceivable that the conjectures for $\kh>0$ are related to the
conjecture for $\kh<0$ through the semi-infinite cohomology of affine
KM algebras. \\
The conjectures naturally fit in a circle of ideas generally referred
to as Kazhdan-Lusztig theory.  This theory interrelates many different
problems, such as the classification of primitive ideals in enveloping
algebras, the computation of the multiplicities in composition series
and the intersection cohomology of Schubert varieties (see~\cite{shi}
for an overview). It applies in particular to simple Lie algebras,
affine KM algebras and quantum groups. In the next section we show
that it also applies to $\W$ algebras.

\section{The KL conjectures for $\W$ algebras}
Compared to the situation for affine Kac-Moody algebras, relatively
little is known about the representation theory of $\W$ algebras. The
fact that a classification of such algebras is still lacking makes it
harder to give a general approach to this problem. We claim, however,
that for the class of $\W$ algebras obtained through hamiltonian
reduction of affine KM algebras, the analogue of most results
described in the previous section exists. In particular, we formulate
the KL conjecture for such $\W$ algebras. 
\subsection{Some generalities on $\W$ algebras and modules}
Let us first consider a general $\W$ algebra, generated by the modes of
a finite set of quasiprimary fields (for a precise definition see~\cite{BS}).  
The $\W$ algebra will have a CSA $h$, \ie a maximal abelian subalgebra of
the zero modes. Unlike for KM algebras, the adjoint action of $h$
on the generators of the $\W$ algebra is in general not diagonalizable
(\eg the zero-mode $W_0$ of the spin 3 field of $\W_3$); therefore the
`triangular' decomposition of $\W=\W_+\oplus h\oplus\W_-$ in positive
root generators, Cartan subalgebra and negative root generators is
given with respect to a subalgebra $h'\subset h$ 
\be
\W = \oplus_{a'} \W_{-a'} \oplus h \oplus_{a'} \W_{a'},
\ee
where $a'$ runs over the set of positive roots $\Delta'_+$. By assumption
$\W_{-a'} \cong \W_{a'}$ as vector spaces, paired by an involutive map
$\sigma : \W_{-a'} \rightarrow \W_{a'}$.\\  
The set-up of representation theory is similar to that of affine KM
algebras, in the following sense. A singular vector $v_a$ is an
eigenvector of the generators of $h$ with weight $a \in h^*$, and
$v_a$ is annihilated by all positive root generators. A highest weight
module $V$ is generated from $v_a$ by the action of the negative root
generators. 
Similarly, one introduces a category ${\cal O}$, which consists of
modules $V$ which have a weight space decomposition into
a direct sum of weight spaces of the subalgebra $h'$, 
\be \label{eq:wdeco}
V = \oplus_{b' \leq a'} V_{b'},
\ee
where the sum is over weights $b'$ satisfying $b' \leq a'$ for $a'$ in
some finite subset of $h'^*$, and $\dim V_{b'} < \infty$ (note that
$b' \leq a'$ iff $a' - b'$ is on the positive root lattice $\Qa'_+$ of $\ha'$).  \\  
The category ${\cal O}$ again contains Verma modules $M(a)$, irreducible quotients
$L(a)$, submodules, etc (but no tensor products as in general the tensor product of
two $\W$ modules is not a $\W$ module).\\  
For every module $V$ in ${\cal O}$ one
can define a (formal) character $\ch V$,  
\be
\ch V = \sum_{b'} \dim V_{b'} e^{b'}.
\ee
The Verma module $M(a)$ has character formula
\be
\ch M(a) = e^{a'} \sum_{b' \in \Qa'_+} P(b') e^{-b'} = e^{a'}\prod_{b' \in
  \Delta'_+} (1-e^{-b'})^{-\dim \W_{b'}},
\ee
where $P(b')$ is some generalized Kostant partition function. \\
The finite dimensionality of the weight spaces $V_{b'}$ implies that
the action of the generators of the CSA $h$ outside $h'$ is reasonably
well-behaved: every weight space $V_{b'}$ can be decomposed into a
finite number of Jordan blocks $U_b$,
\be
V_{b'} = \oplus_b U_b.
\ee
This implies that one can make local composition series in ${\cal O}$,
where the irreducible quotients are again the highest weight modules
$L(b)$, occurring with multiplicities $[V\!:\!L(b)]$. This leads to
character formulas 
$
\ch V = \sum_b [V\!:\!L(b)] \ch L(b),
$
where of course $b$ can only appear in the sum if $b'$ is a weight of
$V$. This applies in particular to a Verma module $M(a)$, leading to 
\be\label{eq:wverrep} 
\ch M(a) = \sum_{b' \leq a'} [M(a)\!:\!L(b)] \ch L(b),
\ee
where clearly $[M(a)\!:\!L(a)] = 1$. 
Once again, this character formula can be inverted, such that the
characters of irreducible modules can be expressed in characters of Verma modules 
\be\label{eq:wirrep}
\ch L(a) = \sum_{b' \leq a'} (L(a)\!:\!M(b)) \ch M(b).
\ee
To conclude: also for $\W$ algebras, the general strategy to find
character formulas is to compute the multiplicities $[M(a)\!:\!L(b)]$.
This is what we will do in the next section.

\subsection{$\W$ modules from $sl_2$ reductions}
A large class of $\W$ algebras can be obtained through a procedure of
(quantum) Hamiltonian reduction of affine KM algebras~\cite{Fea}. A
particularly nice set of reductions are those related to $sl_2$
embeddings~\cite{BTV}. For every $sl_2$ embedding into the simple Lie
algebra underlying the untwisted affine KM algebra, one can define a
BRST complex such that the associated cohomology is non-vanishing only
in the zero-th term. This cohomology is a $\W$
algebra~\cite{FeFr,FKW,BT}. \\ 
Similarly, on the level of the representation theory, the cohomology
of a complex associated to a KM module gives a $\W$ module. This
defines a functor from the category of KM modules to the category of
$\W$ modules. We assume the following properties of this reduction
functor~\cite{FKW,DV}: 
(1)
the cohomology of the BRST complex associated to the KM module is
non-vanishing only in the zero-th term, 
(2)
KM Verma modules $M(\lambda)$ are mapped to $\W$ Verma modules
$M(a(\lambda))$, and 
(3)
a local composition series of a KM Verma module is mapped to a local composition series of the
corresponding $\W$ Verma module.

From these assumptions it immediately follows that, when acting on KM
irreducible modules, the reduction functor maps
\be\label{eq:par}
L(\lambda)\rightarrow L(a(\lambda)) \quad\mbox{or}\quad  L(\lambda)\rightarrow 0,
\ee
where at least one of the maps is to be non-trivial.
If one knows which $L(\lambda)$ have vanishing or
non-vanishing cohomology, then the multiplicities of $\W$ Verma modules
are determined. The main result of this paper is an explicit
formula for these multiplicities, in terms of KL polynomials
associated to a double coset which is completely fixed by the
reduction data. Note that the reduction only gives rise to a $\W$
algebra for $\kh\neq 0$, \ie precisely those weights for which the
KM multiplicities are given by the KL conjecture. This implies that one
has the complete KL conjecture for this class of $\W$ algebras, so
that the characters of all irreducible highest weight $\W$ modules are known. 
    
Let us explain how this should work. Associated to the particular
$sl_2$-reduction is a regular subalgebra $g_r$ of the
finite-dimensional simple Lie algebra $\gf$ underlying the affine
Kac-Moody algebra $\ga$~\cite{DV}. The $sl_2$ subalgebra is principally embedded
into $g_r$. This embedding determines a set of constraints which can
be chosen in such a way that they involve only positive roots. This is
necessary to get non-vanishing cohomology from KM Verma modules. In
explicit examples it is possible to verify that this cohomology is
given by a Verma module of the corresponding $\W$
algebra~\cite{FKW,DV}. We assume that this holds in general.
From the results of~\cite{DV} we expect that the parametrization
$a(\lambda)$ of the $\W$ weight is invariant under the shifted action
of the Weyl group $W^r$ of $g_r$ (which is a finite parabolic subgroup
of $\Wa$). More precisely
\be\label{eq:inva}
a(w.\lambda) =  a(\lambda) \quad \mbox{iff} \quad w\in W^r,
\ee
so there is a one-to-one correspondence between the $\W$-weights and
the invariants of the Weyl group $W^r$.
Using this parametrization we will from now on
denote Verma modules and irreducible modules for the $\W$ algebra by
$M^r(\lambda)$ and $L^r(\lambda)$ with $\lambda$ a weight of $\ga$. Up
to $W^r$ invariance, the labelling by $\ga$ weights fixes the $\W$ weights
uniquely. \\
Let $\lambda$ be a dominant weight. From the existence of the
composition series it follows that the set of primitive weights in a
$\W$ Vermas module $M^r(\lambda)$ is contained in the orbit of the double coset
\be\label{eq:dcos}
W^r_{\lambda} \backslash \Wa_{\lambda}/W^0_{\lambda},
\ee
where $W^r_{\lambda}= W^r\cap W_{\lambda}$.
From the embedding property~(\ref{eq:emb}) of KM Verma modules it now
follows that for each weight on this orbit there is an embedding of
$\W$ Verma modules, thus there is a one-to-one correspondence between
primitive weights and weights on the orbit of the double
coset~(\ref{eq:dcos}). 
\\
It is instructive to note the analogy with the translation functor
discussed in section 2.4: the translation functor maps regular KM Verma
modules $M(\lambda)$ to arbitrary KM Verma modules $M(\lambda')$, such
that the relevant cosets $W_{\lambda}$ are mapped to
$W_{\lambda'}/W_{\lambda'}^0$. Similarly, the reduction functor maps
arbitrary KM Verma modules $M(\lambda)$ to arbitrary $\W$ Verma modules
$M^r(\lambda)$, such that the relevant cosets
$W_{\lambda}/W_{\lambda}^0$ are mapped to $W_{\lambda}^r\backslash
W_{\lambda}/W_{\lambda}^0$. Indeed, the derivation of the $\W$
multiplicities from KM multiplicities from this point on goes completely
analogous to the derivation in section 2.4. 
\\
The irreducible $\W$ module $L^r(\mu)$ may arise only as
the cohomology of the KM modules $L(w.\mu)$ with $w\in W^r$. 
Oviously, the cohomology of the associated KM Verma modules
$M(w.\mu)$ are identical. Therefore, the cohomology of $L(\mu)$ must
vanish when there is a $w\in W^r_{\mu}$ such
that $M(w.\mu)\subset M(\mu)$ with $w.\mu\neq\mu$.
On every $W^r_{\mu}$ orbit of $\mu$, only the lowest weight
contributes therefore. \\
It follows that the reduction functor maps
\be\label{eq:maps}
M_w  \rightarrow  M^r_w
,\qquad
L_w  \rightarrow  L^r_w\delta_{w,\wmax}.
\ee
where again $M^r_w=M^r(w.\lambda),L^r_w=L^r(w.\lambda)$ and
$\wmax$ is the maximal representative of $w$ in the double
coset~(\ref{eq:dcos}).\\
Thus we observe that again the way to associate KL polynomials with
the double coset~(\ref{eq:dcos}) is to take maximal representatives.

To summarize, consider the $\W$ algebra associated with the regular
subalgebra $g_r$. Let $\lambda$ be a dominant weight, and let $w,w'
\in W^r_{\lambda} \backslash \Wa_{\lambda}/W^0_{\lambda}$.  
Denote the double coset of $w$ by $[w]$, the minimal representatives
by $\wmin$ and the maximal representative by $\wmax$, and define the
following polynomials 
\be\label{eq:twoside}
\begin{array}{ll}
P^{IJ}_{w,w'} = P_{\wmax,\wmax'},\qquad& 
Q^{IJ}_{w,w'} = Q_{\wmin,\wmin'}.\\
\tilde{P}^{IJ}_{w,w'} = \sum_{x \in [w]} P_{x,\wmin'} \ve_x \ve_{\wmin'}, \qquad&
\tilde{Q}^{IJ}_{w,w'} = \sum_{x \in [w']} Q_{\wmax,x} \ve_{\wmax} \ve_x. 
\end{array}
\ee
\begin{conjecture}[KL conjecture for $\W$ algebras]
The multiplicities in Verma modules are given by the KL
polynomials associated with the double coset~(\ref{eq:dcos})
\be
\label{eq:klw1}
\begin{array}{lll}
\kh>0:\qquad &
[M^r_w:L^r_{w'}] = P^{IJ}_{w,w'}(1),\quad&
(L^r_w:M^r_{w'}) = \tilde{Q}^{IJ}_{w,w'}(1).\\
\kh<0:\qquad &
[M^r_w:L^r_{w'}] = Q^{IJ}_{w',w}(1),\quad&
(L^r_w:M^r_{w'}) = \tilde{P}^{IJ}_{w',w}(1).
\end{array}
\ee
\end{conjecture}
Hence the character formulas for irreducible $\W$ modules is given by
\be
\label{eq:charfor1}
\begin{array}{ll}
\kh>0:\quad\ch L^r_w=\sum_{w'\geq w} \tilde{Q}^{IJ}_{w,w'}(1) \ch M^r_{w'}.\\
\kh<0:\quad\ch L^r_w=\sum_{w'\leq w} \tilde{P}^{IJ}_{w',w}(1) \ch M^r_{w'}.
\end{array}
\ee
\begin{conjecture}
The embedding diagram of Verma modules corresponds to the Hasse diagram of the double
coset~(\ref{eq:dcos})
\be\label{eq:dihow}
\left.\begin{array}{c}
\kh>0:\quad M^r_{w'}\hookrightarrow M^r_w \\
\kh<0:\quad M^r_w\hookrightarrow M^r_{w'}\end{array}\right\} 
\quad \mbox{iff}
\quad w \leq w' \quad \mbox{with $w,w' \in W^r_{\lambda} \backslash
  \Wa_{\lambda}/W^0_{\lambda}$}. 
\ee
\end{conjecture}
Moreover, we expect that also for $\W$ algebras there is just one
singular vector at any given weight. The KL conjecture supports this
as follows. For $\kh<0$ there is an anti-dominant weight, so here the
proof is identical to the case discussed in section 2.3. Barring a
reflection principle for $\W$ algebras, a general proof for $\kh>0$ is
lacking. However, in the examples we studied, the polynomials appear
to have the property that at arbitrary length $\ell(w)$ one can always
find a $w$ such that polynomial $P_{e,w}=1$. This provides an upperbound for the
number of singular vectors at that weight and consequently also at
every primitive weight $w'.\lambda$ for any $w'\leq w$
(see~(\ref{eq:mulincr})). So we expect that also for $\W$ algebras,
$\dim\mbox{Hom}((M^r(\lambda),M^r(\mu))\leq 1$.

\section{Examples and Applications}
In this section we discuss some examples that on the one hand 
provide evidence for the validity of the conjectures, and on the other
are an illustration of their effectiveness for
actual computations. In particular the (explicit) calculation of $\W$
characters for irreducible highest weight modules is now reduced to
combinatorics on the Weyl group of $\ga$. For simplicity we restrict
to $g=sl_N$. If $\lambda=\sum_{i=0}^l\lambda_i\Lambda_i$ is a dominant
weight (where $\Lambda_i$ are the fundamental weights of $sl_N$,
$\la\Lambda_i,\alphav_j\ra=\delta_{ij}$), then the level
$k=\sum_{i=0}^l\lambda_i$, $P^k_+$ is the set of dominant integral
weights of level $k$, and $P^k_{++}$ are the regular weights in
$P_+^k$. The finite part $\lambdab\in\ol{h}^*$ is
$\lambdab=\sum_{i=1}^l\lambda_i\Lambda_i$ and finally $h^{\vee}=N$.

\subsection{Comparison with known results}
The first check is provided by the Virasoro algebra, which is the quantum
Hamiltonian reduction of the affine KM algebra $g=sl_2$. In that case,
it is a straightforward exercise to show that the conjectures agree
with the results of Feigin and Fuchs: (1) the
embedding diagrams are classified by the double cosets of the
reflection subgroups of the affine Weyl group $\hat{a_1}$ (see Table~\ref{FFtab})
and (2) the multiplicities in the characters are given by the
corresponding KL polynomials: $P_{x,y}=Q_{x,y}=1$ for all $x\leq y$.
 {\sl
\begin{table}[hbt]
\begin{center}
\begin{tabular}{c|c} 
 Feigin-Fuchs        & coset                          \\ \hline
 $I$                 & trivial                        \\ \hline
 $II_{\pm}$          & $a_1$                          \\
 $II^0$              & $a_1/a_1$                      \\ \hline
 $III_{\pm}$         & $\hat{a}_1$                    \\
 $III_{\pm}^0$       & $\hat{a}_1/a_1$                \\
 $III_{\pm}^{00}$    & $a_1 \backslash \hat{a}_1/a_1$ 
\end{tabular}
\end{center}
\caption{\sl classification of embedding patterns of the Virasoro
  algebra.}\label{FFtab}
\end{table}
}

A second check is provided by the $\W_N$ minimal models, which are the
quantum Hamiltonian reduction of the affine KM algebra $g=sl_N$ with
respect to the principal $sl_2$ embedding. Consider the dominant weights $\lambda$ with
$W_{\lambda}$ isomorphic to $W$~\cite{KW1}
\be\label{eq:adm}
\lambda+\rho = w(\Lambda^+ - t \Lambda^-),
\ee
where $t=p/p'$ ($p,p'$ relative prime integers), $\Lambda^+\in P_{++}^{p}$,
$\Lambda^-\in P_{++}^{p'-1}$ and $w$ an arbitrary element of the Weyl
group $\Wf$ of $\gf$.  The simple roots of $\Delta^{re}_{\lambda,+}$
are given by $\hat{\alpha}_i = w(\alpha_i) + \Lambda^-_i \delta$. Let
$\hat{s}_i=r_{\hat{\alpha}_i}$, then  
\begin{itemize}
\item[1.]
$W_{\lambda}$ is generated by the simple reflections $\hat{s}_i$
\item[2.]
$W_{\lambda}^0$ is generated by the $\hat{s}_i$ for which $\Lambda^+_i=0$, 
\item[3.]
$W_{\lambda}^r$ is generated by the $\hat{s}_i$ for which $\Lambda^-_i=0$ and
$\alpha_i$ is a simple root of $g_r$.
\end{itemize} 
The $\W_N$ minimal models arise from dominant weights~(\ref{eq:adm})
which have trivial $W_{\lambda}^0$ and $W_{\lambda}^r$, hence
$\Lambda^+-\rho \in P_{++}^{p-N}$ and $\Lambda^- -\rho\in P_{++}^{p'-N}$. 
The multiplicities $Q_{e,w}$ for these regular dominant are easily read off
from the recusion relation~(\ref{eq:qrecrel2}) for $x=e$, since in
that case $Q_{e,ys} = Q_{e,y}$ for all $sy>y$ so it easily follows
that $Q_{e,w}=1$.  
This reproduces the character formulas of~\cite{FL,FKW} for the
$\W_N$ minimal models. Similarly, admissible modules for arbitrary $\W$
algebras can be obtained. As should be clear from above, the only difference
will be in the domain of $\Lambda^-$.\\ 
New testcases for weights $\lambda$ with $W_{\lambda}\cong W$ arise
when one considers nontrivial subgroups $W_{\lambda}^0$ and/or
$W_{\lambda}^r$ and non-dominant highest weights. In the next section
we will do this for the case of the $\W_3$ algebra. 

\subsection{Classification of $\W_3$ modules}
The Zamolodchikov algebra $\W_3$~\cite{Zam} is the quantum Hamiltonian
reduction of the affine KM algebra $sl_3$ with respect to the
principal $sl_2$ subalgebra~\cite{BO}. The $W_3$ weights are
$(h,w,c)$, the eigenvalues of the zero modes $L_0,W_0$ and $c$
respectively. The parametrization of the $\W_3$ weights in terms of the $sl_3$ weight
$\lambda$ that follows from the BRST construction is 
\be\label{eq:w3norm}
h=\frac{1}{2t}|\lambdab+\rhob|^2 + \frac{c-2}{24},\quad
w=\frac{1}{27t(t-1)}(\lambdab+\rhob,\Lambda_1)
(\lambdab+\rhob,\Lambda_2)
(\lambdab+\rhob,\Lambda_1-\Lambda_2),
\ee
with $c=50-24t-24/t$ for $t=k+3$. The character of a $\W_3$ Verma
module is given by $\ch M^r(\lambda)=q^h \eta(q)^2$ (note that if
$t=1$, the parametrization (\ref{eq:w3norm}) is singular, in that case
one replaces $w\rightarrow (t-1) w$).  \\
Up to isomorphism, there are two nontrivial parabolic subgroups of the
affine Weyl group $\hat{a}_2$ of $sl_3$, namely $a_1\simeq Z_2$ and
$a_2\simeq D_3$. This gives
rise to 9 inequivalent double cosets (we eliminated the
invariance of the $\W_3$ weights under $(t,\lambdab) \rightarrow (1/t,-\lambdab/t)$, which
interchanges $W_{\lambda}^0$ and $W_{\lambda}^r$), see table 2.
For each of these double cosets, we computed the multiplicities
$P^{IJ}_{w,w'}(1)$ (for the first $15$ elements) from the KL
polynomials $P_{w,w'}(q)$ of $sl_3$.
Together with the associated Hasse diagrams, they are given in
tables 5-13. For completeness, we also give the KL polynomials
$P_{w,w'}(q)$ and $Q_{w,w'}(q)$ for $sl_3$, which we computed up till
$\ell(w)=15$, tables 3~\footnote{We thank M.\ Goresky, who first
  computed this table for us.} and 4. \\
To check if the polynomials and Hasse diagrams correspond with
multiplicities and embedding diagrams of $\W$ Verma modules, we
subsequently calculated (parts of) the irreducible characters and
embedding patterns directly on the Verma module.\\
This goes as follows. Starting from a highest weight vector $v_a$
(eigenvector of $L_0,W_0,c$ with eigenvalues $h,w,c$, and annihilated
by the $L_n,W_n$ for $n>0$) a basis $M(a)_{h+N}$ of the
Verma module at depth N is constructed. The rank of the innerproduct matrix (Shapovalov form)
at depth N gives the dimension of the irreducible character at depth
$N$, and the eigenvectors of $W_0$ in the kernel of $L_1,L_2$ and $W_1$ gives the singular vectors. \\
In practice, even at modest depth these calculations require a lot of computer
time: with the Mathematica routines we had available, the singular
vectors could generically be determined up to depth 9, and the
characters up to depth 6\footnote{to appreciate the effectiveness
  of the KL conjecture: applying table 3 to the vacuum module
  for $c=0$ the characters are already determined beyond depth 200,
  where the Verma module has of the order of $10^{20}$ states.}
To verify the predictions of the
conjectures, we selected a dominant weight with every coset  such that
the singular vectors in the associated Verma module occur at the
lowest possible levels, see table 2. For every coset we computed
the characters of the first 15 submodules, and reconstructed the
embedding patterns. We found complete agreement with the KL conjecture. 
\begin{table}[bht]\label{tab3}
\begin{center}
\begin{tabular}{cccrrrrrc}\hline
$W^r_{\lambda}$&$W^0_{\lambda}$ &$t$&$\Lambda^+$&$\Lambda^-$& $c$ &
$h$ & $w$ & {\small Table}\\ \hline
$a_2$&$a_2$ &1   & (1, 0, 0) & (1, 0, 0) &  2 & 0 & $0$ &5 \\
$a_2$&$a_2'$&1   & (0, 1, 0) & (1, 0, 0) &  2 &$\frac{1}{3}$& $\frac{2}{27}$&6 \\
$a_2$&$a_1$ &2   & (1, 1, 0) & (1, 0, 0) &$-10$&$-\frac{1}{3}$& $\frac{1}{27}$&7\\
$a_2$&$a_1'$&2   & (0, 1, 1) & (1, 0, 0) &$-10$& 0&0&8\\
$a_2$& --   &3   & (1, 1, 1) & (1, 0, 0) &$-30$&$ -1$& 0&9\\
$a_1$&$a_1$ &3/2 & (2, 1, 0) & (1, 1, 0) &$-2 $&$-\frac{1}{9}$& $-\frac{1}{81}$&10\\
$a_1$&$a_1'$&3/2 & (2, 0, 1) & (1, 1, 0) &$-2 $&$\frac{2}{9}$& $\frac{10}{81}$&11\\
$a_1$& --   &3/2 & (1, 1, 1) & (1, 1, 0) &$-2 $& 0& 0&12\\
 --  & --   &4/3 & (2, 1, 1) & (1, 1, 1) & 0 & 0 & 0&13\\  \hline 
\end{tabular}
\caption{\sl Classification of $\W_3$ modules with $W_{\lambda}$
  isomorphic to $\hat{a}_2$.}
\end{center}
\end{table}
\subsection{Closed character formulae}
The practical upshot of the KL conjectures is that characters of $\W$
algebras can be computed using only the combinatorics of double cosets
of affine Weyl groups. In general however, the word problem posed by the
recursion relation~(\ref{eq:qrecrel2}) is too complicated to solve in closed form.  
Only in certain special cases, one does get a closed
expression for character formulas, as in the case of the Virasoro
algebra ($Q_{x,y}=1$ for $x\leq y$) and the $\W_N$ minimal models
($Q_{e,y}=1$). We end this section by
discussing two more examples where a closed formula can be obtained:
cosets of type $\Wf\backslash \Wa$ and of type $\Wf\backslash \Wa/\Wf$

\subsubsection{The coset $\Wf\backslash \Wa$}
Consider a coset of the form $\Wf\backslash \Wa$, with
$\Wa$ the affine Weyl group and $\Wf$ the corresponding finite Weyl
group. These cosets correspond to modules on the boundary of
the Kac-table (type $III^0$ of the Virasoro), where the characters are
given by finite sums over Verma characters. \\
For the $\W_3$ algebra $W=\hat{a}_2$ and $\ol{W}=a_2$. The KL
polynomials $P^{IJ}_{x,y}$ and Hasse diagram of $a_2\backslash\hat{a}_2$ is given in
table 9. In that case it can be shown that there are 2
different character formulas, depending only on the length
of $\wmin$ (the minimal representative of $w$ in the coset).  

If the length $\ell(\wmin)$ is odd, there are precisely 2 adjacent
elements of $\wmin$ of length $\ell(\wmin)+1$, of the form $\wmin.j$ and $\wmin.k$
(where $j$ and $k$ are distinct simple reflections). Let $a_2$ denote the $a_2$ generated
by $j$ and $k$. Then the character reads
\be\label{eq:sglhex}
\ch L(w.\lambda) = \sum_{x\in a_2} \epsilon_x \ch M(w.x.\lambda)
\ee
If the length $\ell(\wmin)$ is even, there are at most 3 adjacent
elements of $\wmin$ of length $\ell(\wmin)+1$, and there is precisely one of the form
$\wmin.i$ (for $i$ a simple reflection).
Then we find
\be\label{eq:dblhex}
\ch L(w.\lambda) = \sum_{x\in a_2} \epsilon_x ( \ch M(w.i.x.i.\lambda)
- \ch M(w.i.x.\lambda))
\ee
where again the $a_2$ is generated by $j,k$ ($i,j,k$ are distinct simple reflections).

It appears that these two formulas are related to the generic
decomposition patterns of Weyl modules, studied in~\cite{Lus4}. Similarly,
there is 1 formula in the case of $A_1$, 4 different formulas for
$B_2$ and 12 for $G_2$. \\  
These character formulas apply in particular to the (p,1) topological minimal
models. The admissible weights~(\ref{eq:adm}) in that case are
integral, hence $W^r_{\lambda}=\Wf$. The regular integral weights
\be\label{eq:primfix}
(\lambda+\rho,\alphav_i)\neq 0\bmod p
\ee
are on the orbit of the coset $\ol{W}\backslash W$, with
dominant weights $\lambda\in P^{p}_{++}$. For $\W_3$
therefore, provided $p\geq 3$,~(\ref{eq:sglhex}) and (\ref{eq:dblhex})
give the character formulas for all the regular weights.
We observe that~(\ref{eq:primfix}) is exactly the condition for the
physical states in (p,1) topological minimal matter coupled to
$W$-gravity~\cite{LWS}. An interesting open question is whether the
non-trivial multiplicities indicate the presence of extra physical states. 

The character formulas~(\ref{eq:sglhex}) and (\ref{eq:dblhex}) apply in
other cases as well, for instance (i) for regular integral weights on the boundary
of the $(p,p')$ Kac table, (ii) for weights with $\Wa_{\lambda}^0=\Wf$
and $\Wa_{\lambda}^r$ trivial (interchanging left and right
multiplication) and (iii) for the other $g=sl_3$ related $\W$ algebras
($W_3^{(2)}$ and $sl_3$ itself).

\subsubsection{The coset $\Wf\backslash \Wa/ \Wf$} 
Consider a coset of the form $\Wf\backslash \Wa/\Wf$, with
$\Wa$ an affine Weyl group and $\Wf$ a subgroup isomorphic to the finite Weyl
group $\Wf$. These cosets correspond to modules in a corner of the
Kac-table (type $III^{00}$ of the Virasoro), where again the
characters are given by finite sums over Verma characters, but in
addition they are now grouped in $\gf$ multiplets. 
Specifically,  the multiplicities $P^{IJ}_{x,y}(1)$ are related to the
dimension of weight spaces in finite dimensional modules of
the simple Lie algebra $\gf$~\cite{Lus3}. The correspondence is as follows. \\
First consider the case where the embedding of the left- and right
subgroups is the same and given by $\Wf$ (depending on $\ga$ there may
be more ways to embed $\Wf$ in $\Wa$). Since $\Wa=\Wf\cdot \Qf$
(semi-direct product) and in $\Qf$ there is a unique dominant element
$\alpha$ on each $\Wf$ orbit, it follows that there is a one-to-one
correspondence between coset elements $w\in\Wf\backslash \Wa/ \Wf$ and dominant
roots $\bar{\alpha}_{w}\in P_+\cap\ol{Q}$. More generally, if the
embeddings are chosen differently, this correspondence is between
coset elements $w\in\Wf\backslash \Wa/ \Wf'$ and dominant weights
$\lambdab\in P_+\cap\ol{Q}+\Lambda_i$ where
$\Lambda_i$ is the fundamental weight that determines the embedding:
$\Wf'$ is generated by the set of simple reflections $s_j$ for
$j\neq i$. 
Then the result of~\cite{Lus3} states that
\be\label{eq:lus1}
P_{w,w'}^{IJ}(1)=\dim L(\lambdab_{w'})_{\lambdab_w}
\ee
where $L(\lambdab)_{\mub}$ is the weight space of weight $\mub$ in the
(finite dimensional) irreducible $\gf$ module of highest
weight $\lambdab$.   \\
This result applies in particular to the $\W_N$ algebras with
$c=\mbox{rank}({\gf})$, \ie with $\kh=1$. From~(\ref{eq:adm}) it
follows that the dominant weight is determined by a level 1 weight,
\ie $\lambda+\rho=\Lambda_i$. $W_{\lambda}^r=\Wf$ and
$W_{\lambda}^0=\Wf'$. Using the correspondence described
above, to every primitive weight $\mu=w.\lambda$ one
associates the dominant weight $\lambdab_w=\mub+\rhob$. Then the sum
$w'\geq w$ can be rewritten as follows    
\be\label{eq:karc2}
\ch M^r(\mu) = \sum_{\mub'\geq\mub} \dim
L(\mub'+\rhob)_{\mub+\rhob} \ch L^r(\mu').
\ee
Inverting this (using the basis transformation from Verma modules
$M(\mub)$ to singlets $e^{\mub}$) gives 
\be
\label{eq:dblhexc2}
\ch L^r(\mu) = \sum_{x\in\Wf} \epsilon_x \ch M^r(\mu+\rho-x(\rho))
\ee
This character formula was first proposed in~\cite{MMMO}, where it was
obtained as a limit of the characters of the $\W_N$ minimal
models~\cite{FL}. The inverse formula~(\ref{eq:karc2}) was obtained
for $\W_3$ in~\cite{BMP}, using an explicit construction of the
singular vectors in the Fock space and comparison of the characters for each side. 
Again, this character formula applies more generally, in particular to
the weights in a corner of the $(p,p')$ Kac table (in that case one
replaces $\rho\rightarrow p\rho$ everywhere on the RHS
of~(\ref{eq:dblhexc2})) 

\section{Concluding remarks}
In this paper we have formulated the KL conjecture for $\W$ algebras
associated with arbitrary $sl_2$ reductions. The result can be
described in terms of a double coset $W_{\lambda}^r\backslash
W_{\lambda}/W_{\lambda}^0$: the Hasse diagram gives the embedding
diagram of the Verma modules, and the KL polynomials give the
multiplicities in the characters. 

The conjectures also apply to finite $\W$ algebras, which
are the Hamiltonian reduction of simple Lie algebras $\gf$~\cite{BT}.
In that case one simply takes $W$ to be the Weyl group $\Wf$ of $\gf$.
The character formulas for this class of algebras are given in~\cite{DV}
(for regular integral weights only) and the results agree
completely.

We remark that the conjecture is also a useful tool to analyse the
structure of the Verma modules in more detail. This is particularly
important if one attempts to construct a resolution of the irreducible
modules by Verma modules. The physical motivation for doing so is the
application to $\W$ gravity/strings: it is much simpler to compute the
(string) BRST cohomology on Verma modules than on irreducible modules.
The problem is that it is not always possible to find a resolution by
Verma modules. For instance, for the $\W_3$ string at $c=2$ it is found by
explicit construction~\cite{BMP} that there is no resolution by Verma
modules, but instead one is forced to introduce generalized Verma
modules. This is directly linked to the existence of primitive vectors
that are pseudo-singular rather than singular (they are not an
eigenvector of $W_0$). For such an analysis it is convenient to have
the data presented by the KL conjecture. This way for instance, one
can easily show that the character~(\ref{eq:dblhex}) of the (p,1) topological
minimal models for $\W_3$ cannot be reproduced by a resolution by
(generalized) Verma modules. This is due to the occurence of subsingular
vectors~\cite{DV}. It would be very interesting to know what type of
modules are needed to build such a resolution, since these modules are
going to carry the cohomology of the topological $\W_3$ string.
Work on this is in progress.

\renewcommand{\theequation}{A.\arabic{equation}}
\setcounter{equation}{0}

\section*{Appendices}
\subsection*{KL polynomials on Coxeter groups}
In this appendix, we summarize the definition and some of the properties of KL
polynomials $P_{x,y}$ for a Coxeter group $W$, for details
see~\cite{KL1,Hu}. 
The starting point is the Hecke algebra ${\cal H}$ with generators
$T_y$ (one for each $y \in W$) and defining relations  
\bea\label{eq:hecke}
T_x T_{y} = T_{xy} & \mbox{if} & \ell(xy) = \ell(x) + \ell(y) \\
(T_s + 1)(T_s-q) = 0 & \mbox{if} & s \in S
\eea
where $S$ is the set of simple reflections that generate $W$. The
elements $T_y$ are invertible in ${\cal H}$, and one can write  
\be\label{eq:hecke-inverse}
T_{y^{-1}}^{-1} = \sum_{x \leq y} \ve_x \ve_y R_{x,y}(q) q^{-\ell(y)} T_x
\ee
where $\ve_y = (-1)^{\ell(y)}$, and $R_{x,y}(q)$ is a polynomial
in $q$ of degree $\ell(y)-\ell(x)$ for $x\leq y$, uniquely defined
by~(\ref{eq:hecke-inverse}). The map $\imath$ defined by 
\be\label{eq:hecke-auto}
\imath (q) = q^{-1},\qquad  \imath (T_y) = T^{-1}_{y^{-1}}
\ee
is an automorphism of ${\cal H}$. The KL polynomials are associated
with the invariants of $\imath$. For any pair $x \leq y$ in $W$, there is a
uniquely defined polynomial $P_{x,y}$ of degree $\leq (\ell(y)
-\ell(x)-1)/2$ if $x < y$, and $P_{x,x}=1$, such that 
\be
C_y = \sum_{x \leq y} \ve_x \ve_y q^{\ell(y)/2-\ell(x)}P_{x,y}(q^{-1})T_x
\ee
satisfies 
\be
\imath(C_y) = C_y
\ee
for all $y \in W$. Equivalently, the $P$-polynomials satisfy
\be
q^{\ell(y)-\ell(x)} P_{x,y}(q^{-1}) = \sum_{x \leq z \leq y} R_{x,z}
P_{z,y}(q) \quad \mbox{for all $x \leq y$} 
\ee
From this, one can extract a recursion relation (expressing the
polynomials $P_{x,y}$ in terms of the polynomials $P_{x',y'}$ with
$y'<y$). Namely, for $ys>y$ one has~\cite{KL1}
\be\label{eq:recrel}
P_{x,ys}=q^{1-c}P_{xs,y}+q^cP_{x,y}-q\sum_{\stackrel{x\leq
    z<y}{zs<z}} P_{x,z}\slash{\!\!\!\!P}_{z,y}
\ee
Here $c=1$ if $xs<x$ and $0$ otherwise, and $\slash{\!\!\!\!P}_{z,y}$ is the 
term in $P_{z,y}$ of (maximal) degree $\frac{1}{2}(\ell(y)-\ell(z)-1)$. 
The initial values of the recursion relation are
$P_{x,e}(q)=\delta_{x,e}$. This implies in particular that
$P_{x,y}(q)=0$ unless $x\leq y$. From~(\ref{eq:recrel}) it also
follows that $P_{x,y}(0)=1$ if $x\leq y$. In the case crystallographic
Coxeter groups (which includes (affine) Weyl groups)
the coefficients of $P_{x,y}$ give the dimensions of stalks of
cohomology sheaves of the intersection cohomology complexes associated
to Schubert varieties~\cite{KL2}. This implies in particular that
these coefficients are nonnegative integers.   

Similarly, if $ys<y$ it can be shown that $C_y T_s = -C_y$ which
implies that
\be\label{eq:pshift}
P_{x,y} = P_{xs,y} \quad \mbox{for $x \leq y$ and $ys<y$}
\ee 
For finite Coxeter groups (where there is a unique longest element
$w_0$) it easily follows that 
\be
P_{x,w_0}=1
\ee 

\subsection*{Inverse KL polynomials on Coxeter groups}
The KL polynomials $P_{y,w}$ form an upper triangular
matrix with 1's on the main diagonal, which naturally can be inverted.
Thus, one can define for each $x \leq y$ in $W$ a polynomial $Q_{x,y}$
such that  
\be\label{eq:invdef}
\sum_{x \leq z \leq y} \ve_x \ve_z P_{x,z} Q_{z,y} =
\delta_{x,y} \quad \mbox{for all $x \leq y$} 
\ee
It is clear that $Q_{x,x}(q)=1$ and that $Q_{x,y}(q)$ has degree $\leq
(\ell(y)-\ell(x)-1)/2$ for $x<y$, and $\slash{\!\!\!\!Q}_{x,y} =
\slash{\!\!\!\!P}_{x,y}$. It also follows that  
\be
q^{\ell(y)-\ell(x)} Q_{x,y}(q^{-1}) = \sum_{x \leq z \leq y} Q_{x,z}
R_{z,y}(q) \quad \mbox{for all $x \leq y$} 
\ee
The $Q$-polynomials are associated to invariants of $\imath$. Define
elements $S_x,D_x$ of ${\cal H}^*$ by   
\be
\langle S_x, \imath(T_y) \rangle = \langle D_x , C_y \rangle = \delta_{x,y^{-1}}
\ee
and let $\langle \imath(u), h \rangle = \imath( \langle u, \imath(h)
\rangle)$. It follows that $\imath (D_x) = D_x$ and 
\be
D_x = \sum_{x \geq y} q^{\ell(x)/2 - \ell(y)} Q_{x,y} S_y
\ee
with $Q_{x,y}$ the inverse polynomials~(\ref{eq:invdef}). From the
right ${\cal H}$-action on ${\cal H}^*$ given by 
$\langle u \cdot h,h' \rangle = \langle u, h h' \rangle$
one can conclude that $D_x \cdot T_s = q D_x$ if $xs>x$. This implies 
\be\label{eq:qshift}
Q_{x,ys} = Q_{x,y} \quad \mbox{for $x \leq y$ and $xs>x$}
\ee
From this it easily follows that
\be\label{eq:nulfix}
Q_{e,y} = 1 \quad \mbox{for all $y \in W$}
\ee
The analogue of~(\ref{eq:recrel}) for the $Q$-polynomials is (for $xs<x$)
\be\label{eq:qoldrec}
Q_{xs,y}=q^{1-c}Q_{x,ys}+q^cQ_{x,y}-q\sum_{\stackrel{x <
    z \leq y}{zs>z}} \slash{\!\!\!\!Q}_{x,z} Q_{z,y}
\ee
with $c=1$ for $ys>y$ and $c=0$ for $ys<y$. 
In this form~(\ref{eq:qoldrec}) cannot be used to solve for the $Q$-polynomials
(at least not in the case of infinite Coxeter groups). But,
combining~(\ref{eq:qoldrec}) and~(\ref{eq:qshift}) for $ys>y$, one
gets a useful relation (expressing $Q_{z,w}$ in terms of $Q_{z',w'}$
with $w'<w$) 
\be\label{eq:qrecrel}
Q_{x,ys} = c Q_{xs,y} + (-q)^c Q_{x,y} + c q \sum_{\stackrel{x <
    z \leq y}{zs>z}} \slash{\!\!\!\!Q}_{x,z} Q_{z,y}
\ee
with $c=1$ for $xs<x$ and $c=0$ for $xs>x$. 
In the case of (affine) Weyl groups the coefficients of $Q_{x,y}$ are again
nonnegative integers. \\
In the case of a finite Coxeter group, the KL polynomials $P$ are
related to the inverse polynomials $Q$ through 
\be
Q_{x,y} = P_{w_0y,w_0x}
\ee
On the other hand: if the group is not finite, $P$ and $Q$ do not
appear to be related in any such way.

\subsection*{KL polynomials on cosets}
Similar as for ordinary Coxeter groups $W$, one can also construct KL
polynomials on cosets $W/W_I$ (or $W_I\backslash W$), where
$W_I$ is a parabolic subgroup of $W$~\cite{Deo}. In particular one
finds that again there are recursion relation for the associated
polynomials $P^I_{x,y}$ and $Q^I_{x,y}$. If the subgroup $W_I$ is 
finite (which is sufficient for our purposes) the polynomials
$P^I,Q^I$ and their inverses $\tilde{Q}^I,\tilde{P}^I$ defined
through\footnote{Note that this notation is slighly different from the
  notation used in~\cite{DV}.}
\be
\sum_{x \leq z \leq y} \tilde{Q}^{I}_{x,z} P^{I}_{z,y} = \sum_{x \leq
  z \leq y} Q^{I}_{x,z} \tilde{P}^{I}_{z,y} = \delta_{x,y} 
\ee
can be expressed in terms of the KL polynomials of $W$ 
\be\label{eq:oneside2}
\begin{array}{ll}
P^{I}_{x,y} = P_{\xmax,\ymax} & 
Q^{I}_{x,y} = Q_{\xmin,\ymin} \\
\tilde{P}^{I}_{x,y} = \sum_{z \in [x]} P_{z,\ymin} \ve_z \ve_{\ymin}\qquad &
\tilde{Q}^{I}_{x,y} = \sum_{z \in [y]} Q_{\xmax,z} \ve_{\xmax} \ve_z 
\end{array}
\ee
Here  $\zmin$ is the minimal- and
$\zmax$ is the maximal representative of the coset $[z]$ of $z$. 

However, the cosets that play a role in this paper are two-sided cosets 
$W_I\backslash W / W_J$, with respect to parabolic subgroups 
$W_I$ and $W_J$. There does not appear to be an abstract
set-up for double-sided cosets. In particular, the partial ordering on
these cosets is more complicated than in the case of one-sided cosets
(for example, the length of adjacent elements may differ by more than
1). So instead of defining KL polynomials through a recursuion
relation we take~(\ref{eq:oneside2}) as our starting point, \ie we
define
\be
\begin{array}{ll}
P^{IJ}_{w,w'} = P_{\wmax,\wmax'} & 
Q^{IJ}_{w,w'} = Q_{\wmin,\wmin'} \\
\tilde{P}^{IJ}_{w,w'} = \sum_{x \in [w]} P_{x,\wmin'} \ve_x \ve_{\wmin'} \qquad &
\tilde{Q}^{IJ}_{w,w'} = \sum_{x \in [w']} Q_{\wmax,x} \ve_{\wmax} \ve_x 
\end{array}
\ee
where the polynomials $\tilde{P}^{IJ},\tilde{Q}^{IJ}$ are again the
inverse polynomials
\be
\sum_{x \leq z \leq y} \tilde{Q}^{IJ}_{x,z} P^{IJ}_{z,y} = \sum_{x
  \leq z \leq y} Q^{IJ}_{x,z} \tilde{P}^{IJ}_{z,y} = \delta_{x,y} 
\ee

\section{Tables}
Table 3:
{\sl
KL polynomials $P_{x,y}$ for the Weyl group $\hat{a}_2$ of the affine KM
algebra $g=sl_3$, up to $\ell(y)=15$. To find $P_{x,y}$ for arbitrary
pairs $x,y$ (with $\ell(y)\leq 15$) use that\\
1. $P_{x,y}=0$ unless $x\leq y$,  \\
2. $P_{x,y}=P_{x^{-1},y^{-1}}$, \\
3. $P_{x,y}=P_{\tau(x),\tau(y)}$ with $\tau$ an automorphism of the
  Dynkin diagram, \\
4. $P_{x,y}=P_{x',y}$ for $x\leq x'$ and $P_{x',y}(1)$ maximal, \\ 
5. if 1-4 does not apply then $P_{x,y}=1$.\\
So, given a pair $x,y$ with $x\leq y$, one first fixes i,j,k and an
order (\ie reading from left-to-right or from right-to-left) such that
$y$ is in the table.  Secondly, one searches for an $x'$ (in the fixed
order) in the table such that $x\leq x'$ and $P_{x',y}(1)$ is maximal;
then $P_{x,y}=P_{x',y}$. If either of the two steps fail then
$P_{x,y}=1$.  
}\\[2mm]
Table 4:
{\sl 
Inverse KL polynomials $Q_{x,y}$ for the Weyl group $\hat{a}_2$ of the affine KM
algebra $g=sl_3$, up to $\ell(x)=14$. To find $Q_{x,y}$ for arbitrary
pairs $x,y$ (with $\ell(x)\leq 14$) use the rules of table 3. with x,y
interchanged and the ordering reversed. 
}

\newpage
\epsfxsize = 14 cm
\epsfbox[100 -100 500 800]{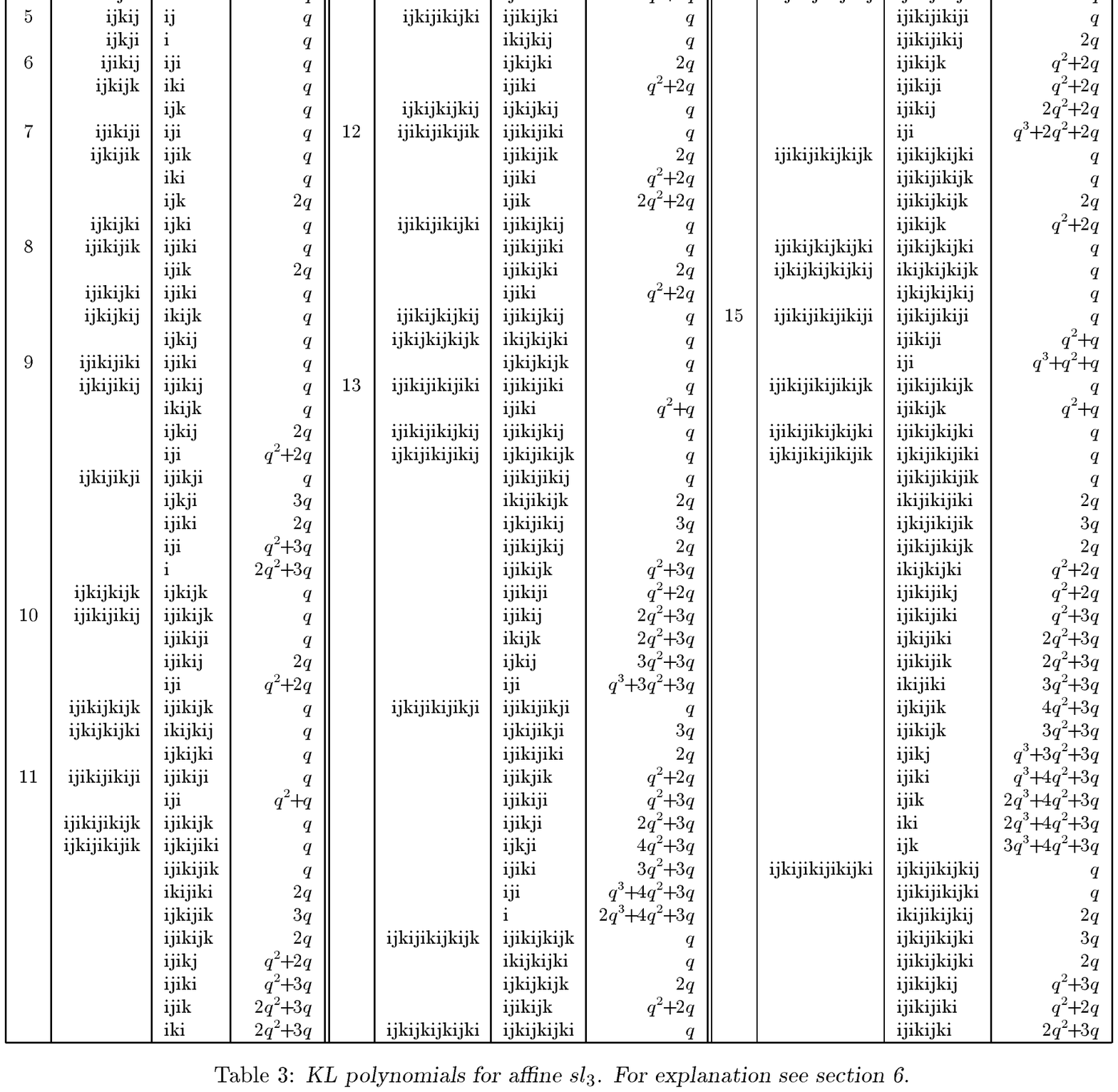}

\newpage
\epsfxsize = 13.5 cm
\epsfbox[100 -100 500 800]{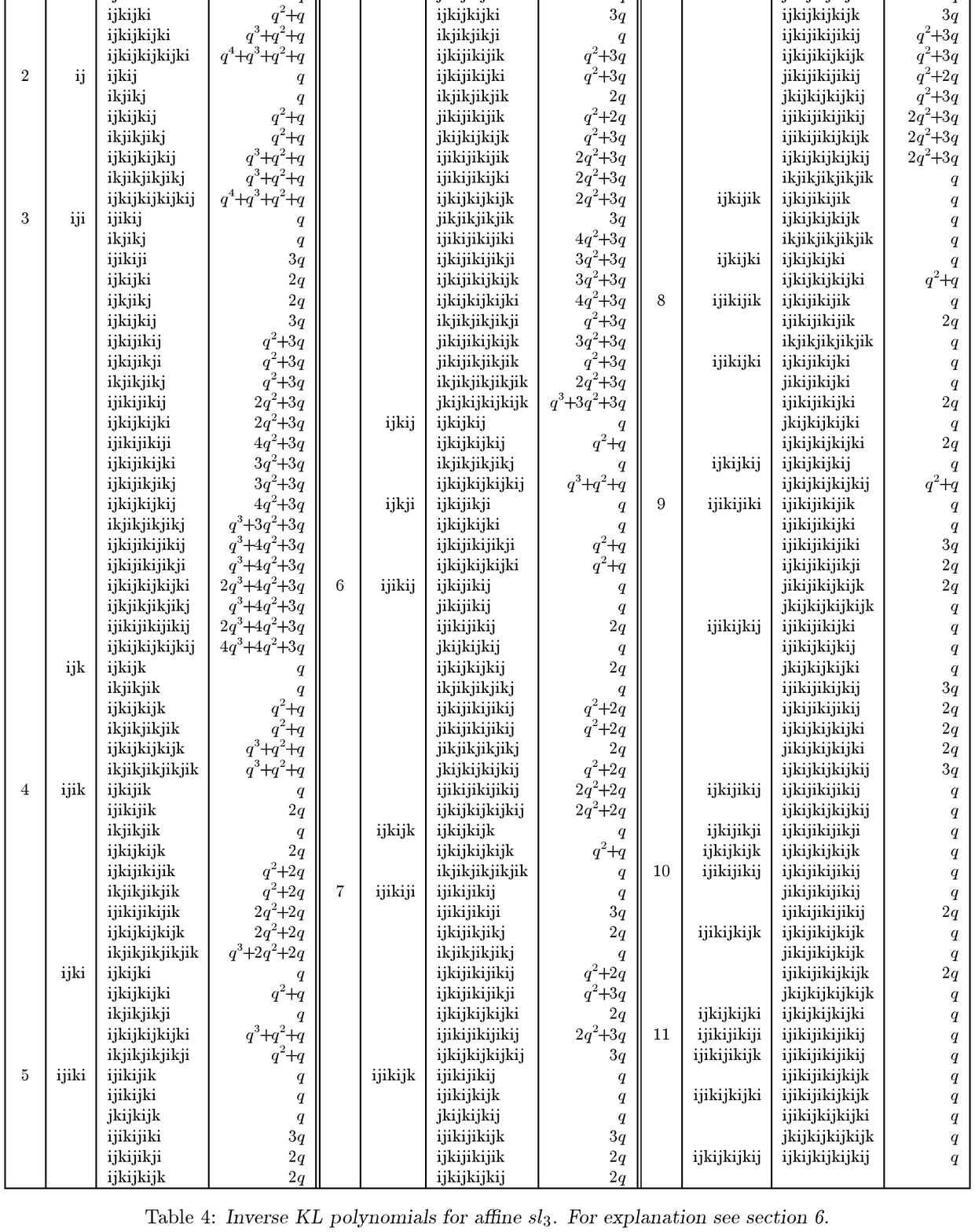}

\newpage
\epsfxsize = 14 cm
\epsfbox[50 -100 450 800]{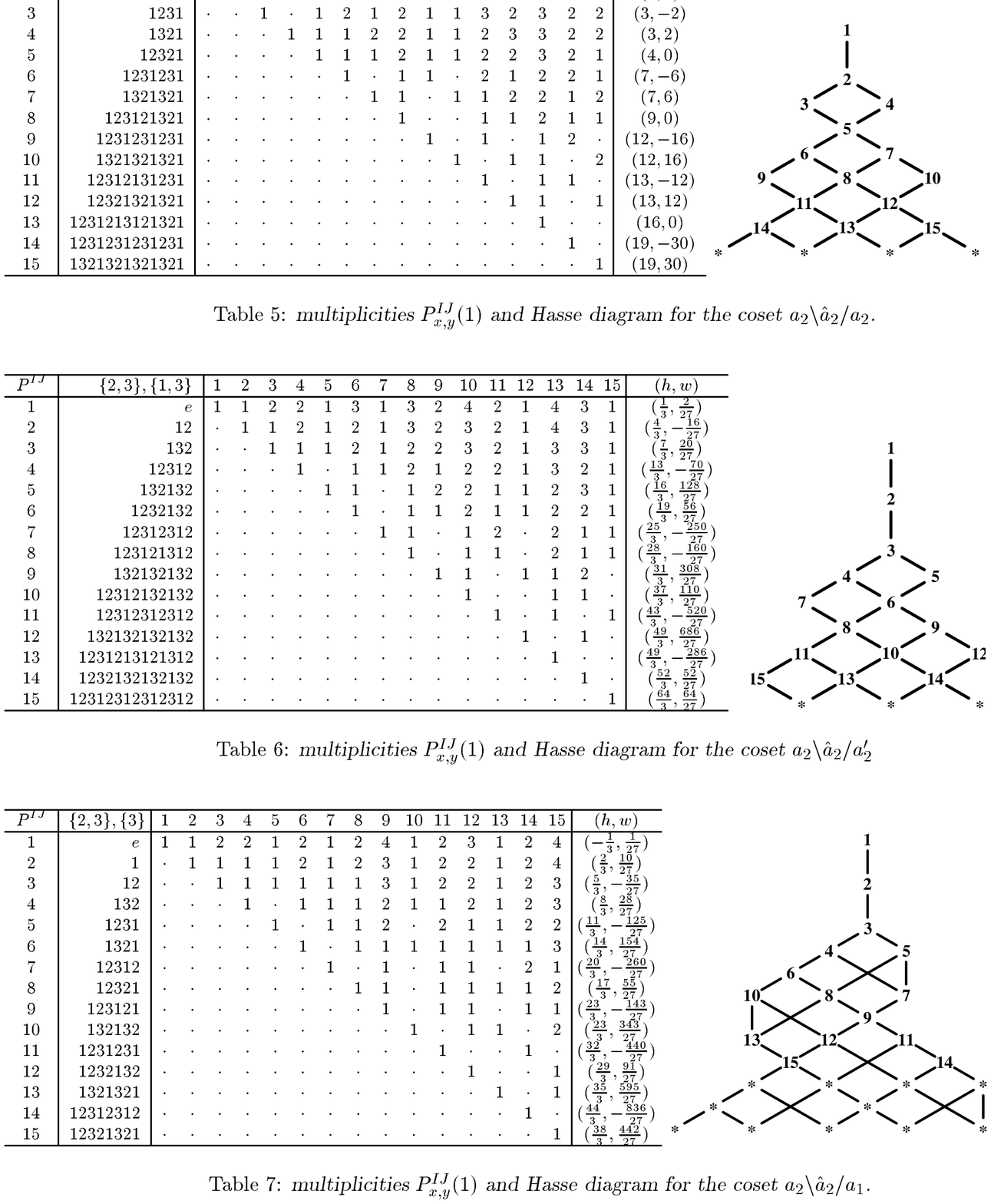} 

\newpage
\epsfxsize = 14 cm
\epsfbox[50 -100 450 800]{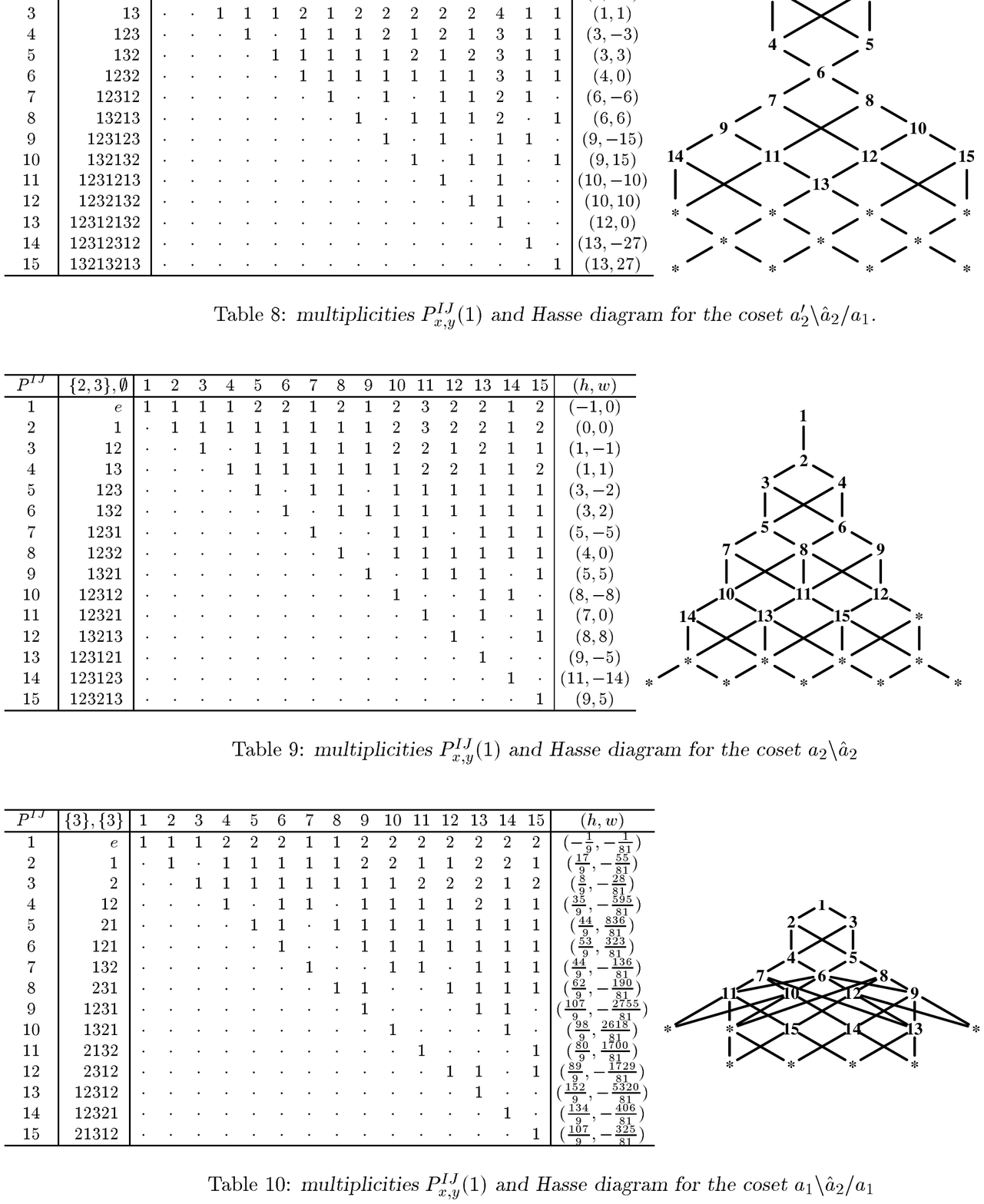}

\newpage
\epsfxsize = 14 cm
\epsfbox[50 -100 450 800]{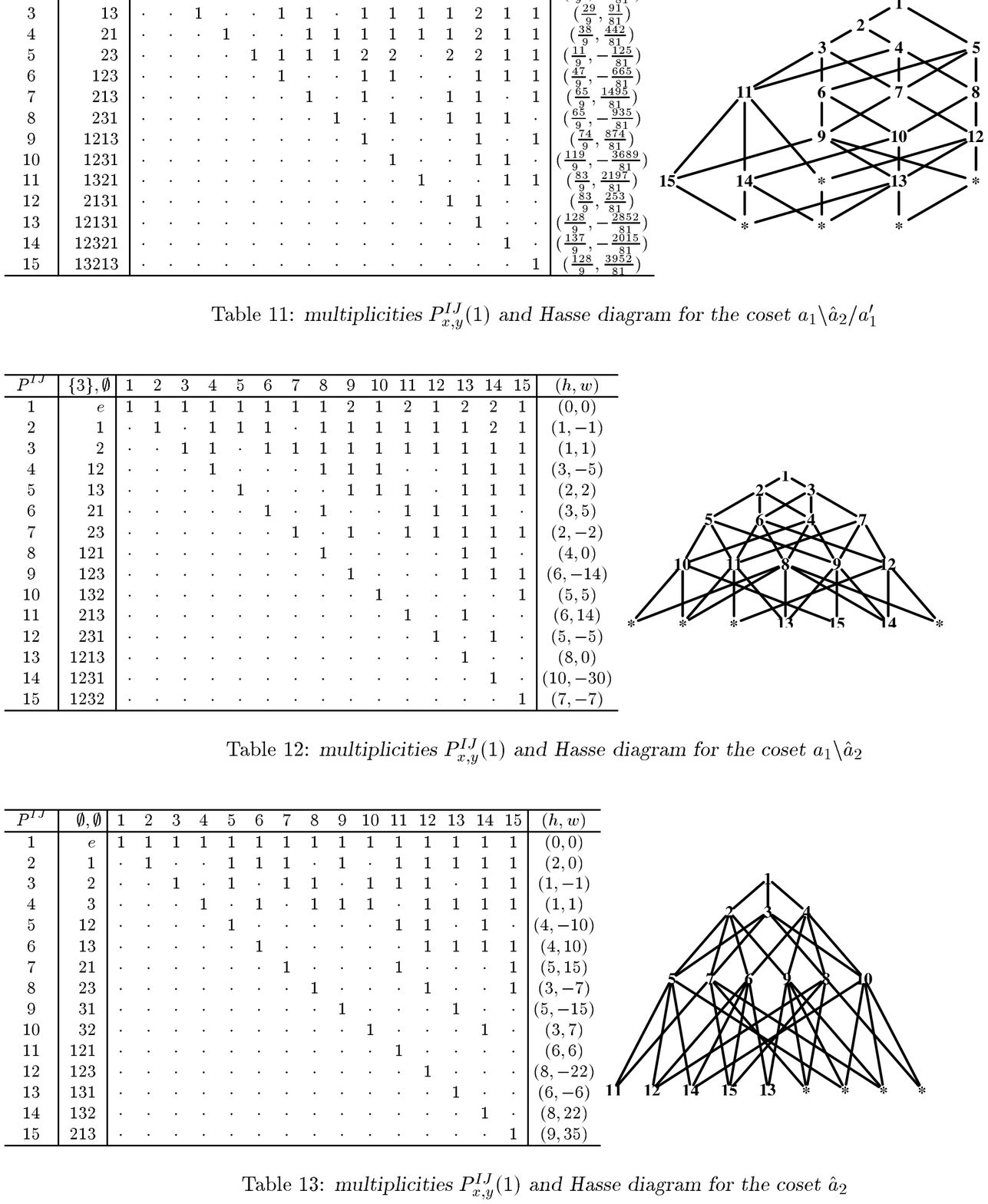}
\newpage

\hoffset = 0 cm
\section*{Acknowledgements}
Discussions with P.\ Bouwknegt, W.\ Lerche, K.\ Pilch, W.\ Soergel, G. Zuckerman
and especially M.\ Goresky are gratefully acknowledged. KdV thanks the
Dept. of Applied Mathematics in Durham for hospitality and support. \\
KdV is supported by the EC Human Capital and Mobility Program.

\end{document}